# Motile *Escherichia coli*-laden Droplets Exhibit Reduced Adhesion and Anomalous Wetting Behavior

Sirshendu Misra[±,a,b], Sudip Shyam[±,a], Priyam Chakraborty[a,c], Sushanta K. Mitra[a],*

[±]: *Authors contributed equally*

*: Corresponding author. Email: skmitra@uwaterloo.ca

[a] Micro & Nano-Scale Transport Laboratory, Waterloo Institute for Nanotechnology, Department of Mechanical and Mechatronics Engineering, University of Waterloo, 200 University Ave W, Waterloo, Ontario N2L 3G1, Canada

[b] Current Affiliation: Department of Mechanical Engineering, Indian Institute of Science, Bengaluru, 560012, India

[c] Current Affiliation: Department of Aerospace Engineering, Indian Institute of Technology Kharagpur, 721302, India

## 1. ABSTRACT

**Hypothesis:**

Bacterial contamination of surfaces poses a major threat to public health. Designing effective antibacterial or self-cleaning surfaces requires understanding how bacteria-laden droplets interact with solid substrates and how readily they can be removed. We hypothesize that bacterial motility critically influences the early-stage surface interaction (i.e., surface adhesion) of bacteria-laden droplets, which cannot be captured by conventional contact angle goniometry.

**Experiments:**

Sessile droplets containing live and dead Escherichia coli (*E. coli*) were studied to probe their wetting and interfacial behavior. Contact angle goniometry was used to probe dynamic wetting, while a cantilever-deflection-based method was used to quantify adhesion. Internal flow dynamics were visualized using micro-particle image velocimetry ($\mu - \text{PIV}$) and analyzed statistically. Complementary sliding experiments on moderately wettable substrates were performed to assess contact line mobility under tilt.

**Findings:**

Despite lower surface tension, droplets containing live *E. coli* exhibited lower surface adhesion forces than their dead counterparts, with adhesion further decreasing at higher bacterial concentrations. μ-PIV revealed that flagellated live *E. coli* actively resist evaporation-driven capillary flow via upstream migration, while at higher concentrations, evidence of collective dynamics emerge, arguably producing spatially coherent bacterial motion despite temporal variability. These coordinated flows disrupt passive transport and promote depinning of the contact line, thereby reducing adhesion. Sliding experiments confirmed enhanced contact line mobility and frequent stick-slip motion in live droplets, even with lower receding contact angles and higher hysteresis. These findings demonstrate that *E. coli* motility plays a critical role in modulating early-stage droplet-surface interactions and provide mechanistic insight into droplet retention and contact-line mobility, the understanding of which is crucial for the removal of contaminated droplets from solid surfaces.

## 2. INTRODUCTION

Bacterial contamination of surfaces has significant implications for both public health and industrial operations. Surface-deposited droplets carrying pathogenic bacteria are major vectors for fomite-based transmission[1], particularly in healthcare settings and public spaces. Upon drying, these droplets can leave behind viable bacteria, leading to persistent contamination and increased risk of infection. Prolonged residence of bacterial droplets on such surfaces promotes irreversible attachment[2] and facilitates biofilm[3–5] formation, which is notoriously difficult to remove and contributes to persistent infections[6] in both medical and industrial settings. As a result, the ability to promptly and effectively remove bacteria-laden droplets from solid substrates is critical for limiting surface-mediated transmission and designing hygienic, self-cleaning, or antibacterial surfaces[7].

Much of the existing literature on bacterial adhesion has focused on the long-term attachment of individual cells or microbial communities onto solid surfaces, governed by thermodynamic (e.g., Lifshitz-van der Waals, acid-base, and electrostatic interactions, as described by the classical Derjaguin-Landau-Verwey-Overbeek (DLVO) and extended-DLVO theories[8,9]) and kinetic factors[10–12], alongside the influence of surface functionalization/chemistry[13,14], surface topography and roughness[15–17], and secreted extracellular polymeric substances (EPS)[18–22]. While such studies have provided deep insights into bacterial colonization and biofilm development, our work focuses on a distinct but related problem: *the adhesion of bacteria-laden droplets, i.e., the force required to detach a droplet from a surface, rather than the attachment of bacteria themselves*. While some recent studies explored droplet-scale interactions of bacteria with surfaces, such as the study by Recupido *et al.*[23] on forced wetting of evaporating bacterial droplets, the work of Sempels *et al.*[24] on biosurfactant-mediated reversal of the coffee-ring effect, and Agrawal et al.[25] on bacterial deposition dynamics in drying drops, these studies primarily focus on evaporation-driven deposition and contact line behavior. However, a clear mechanistic understanding of how bacterial motility modulates early-stage droplet adhesion forces remains relatively unexplored[26]. Understanding this interfacial interaction is critical from a practical standpoint, as the ease of removing contaminated droplets directly affects the efficacy of surface cleaning and pathogen mitigation. Our study aims to address this gap.

A key factor in addressing this problem is a mechanistic understanding of how bacterial motility modifies the internal dynamics and near-surface behavior of sessile droplets. In a pinned droplet, evaporation-driven capillary flow[24,27] typically governs internal transport. However, the presence of self-propelled, motile bacteria within the droplet can introduce competing effects that can locally resist or reorganize this flow field. This interplay is expected to reconfigure internal circulation, alter contact line dynamics, and ultimately modify the droplet-surface adhesion force.

We use *Escherichia coli* (*E. coli*), a gram-negative, flagellated bacterium commonly found in the human intestine, as a model pathogen. It is widely used[28] in research due to its ease of cultivation and handling, extensive genetic data, and well-documented taxonomy. *E. coli* typically exhibits helical propulsion, where its flagellum (or flagellar bundle) rotates as a helix, resulting in forward motion. Its trajectory resembles a random walk, commonly referred to as “run and tumble” dynamics[29,30]. During the run phase, the bacterium moves in a straight line, propelled by the counterclockwise rotation of the flagella, while in the tumble phase, a clockwise rotation of at least one flagellum causes directional changes. These motile behaviors are of critical importance in bacteria-laden sessile droplets, where the bacterial population can respond to internal flow fields and external stimuli through various complex taxis mechanisms, such as chemotaxis[31,32],
phototaxis[33], pH taxis[34], and rheotaxis[35–39]. These taxis allow bacteria to orient along streamlines
and exhibit preferential directional motion within the droplet. Rheotaxis, in particular,
characterized by upstream migration in response to shear gradients, has been widely observed in
various motile bacterial species. In sessile, bacteria-laden droplets with capillary convection[24,27],
these taxis behaviors can significantly alter the internal velocity field, adding complexity to the
overall flow behavior. At higher concentrations, motile bacteria can also exhibit collective
dynamics[40–43,26], further altering flow organization and interfacial behavior. Droplets containing
motile bacteria can therefore be viewed as confined active suspensions rather than passive particle-
laden fluids. In active bacterial suspensions[44], individual cells convert chemical energy into
mechanical propulsion, producing self-driven transport, local hydrodynamic disturbances, and
modified transport/rheological behavior under suitable conditions. In a sessile droplet, these effects
occur within a confined interfacial geometry bounded by the solid substrate, the liquid-air
interface, and the three-phase contact line. Prior studies[45,46] have shown that bacterial trajectories
near boundaries can be modified by hydrodynamic interactions, wall alignment, and confinement.
Consequently, bacterial motility, evaporation-driven capillary flow, boundary interactions, and
rheotactic migration may jointly influence the internal motion field and contact-line dynamics of
bacteria-laden droplets.

Thus, the interaction between bacteria-laden droplets and solid surfaces becomes a
multifaceted phenomenon governed by capillarity, motility, taxis, and collective effects. This study
takes a dual-scale approach to unravel the early stage ($\sim O(100\text{s})$) interfacial dynamics of such
motile bacteria-laden droplets, investigating both macroscopic wetting behavior and particle-scale
migration dynamics, using *E. coli* as the model microbe. We compare the surface interactions of
live and dead *E. coli*-laden droplets using complementary experimental techniques. At the droplet
scale, dynamic wetting is characterized through contact angle measurements, while surface
adhesion forces are directly quantified with a cantilever-based method[47–51]. At the particle scale,
micro-particle image velocimetry (μ-PIV) is used to visualize internal flow fields, capturing the
interplay between bacterial motility and bulk capillary flow dynamics.

Our findings reveal distinct differences in wetting and adhesion between live versus dead
*E. coli*-laden droplets, linked to variations in motility and contact line mobility. These insights
provide a mechanistic perspective on how motile bacterial populations modulate droplet retention and surface interaction, with implications for the design of advanced antibacterial and antifouling surfaces.

# 3. MATERIALS AND METHODS

## 3.1. Preparation of Bacterial samples:

We used *Escherichia coli* (K-12 Strain SMG 123, PTA-7555, ATCC, Cedarlane, Canada) for our experiments. Bacterial samples were prepared by adding lyophilized *E. coli* to pre-autoclaved Lauryl Tryptose (LT) broth (catalog no.: DF0241170, Difco). The solution was incubated at 37°C with 100 rpm agitation overnight (approximately 16-18 hours), which corresponds to the stationary phase in the growth curve. Following incubation, 1 ml of the culture was centrifuged at 10,000 rpm for 10 minutes, allowing the *E. coli* cells to form a pellet while the supernatant was removed. The pellet was washed twice with deionized (DI) water to remove residual culture media and then resuspended in 1 ml of fresh DI water. We note that DI water is not a physiological suspending medium and may impose osmotic stress on *E. coli* cells. However, this preparation was selected to minimize contributions from residual culture-media components to surface tension, wetting, and adhesion measurements. Bacterial motility was confirmed before experiments using motility agar assay and fluorescence microscopy. Note that in one of our earlier studies[49], the same qualitative trend in adhesion signature of live/dead *E. coli*-laden droplets was observed across three different suspending media, namely DI water, phosphate-buffered saline, and Lauryl Tryptose media broth, indicating that the observed adhesion trend is not specific to DI water as the suspending medium.

Agar plate culture confirmed that the final stock solution had a bacterial concentration of ~ $10^7$ CFU/ml (colony forming units). Serial dilution were carried out to further attain concentrations of $10^6$ CFU/ml and $10^5$ CFU/ml, respectively. We also prepared dead bacterial samples by suspending the live bacterial pellets in an aqueous solution of 70% isopropyl alcohol (IPA). For dead *E. coli* suspensions, the concentration is reported as the nominal cell-number-equivalent concentration, estimated from CFU count of the corresponding live suspension before IPA treatment, and should not be interpreted as viable CFU/mL. Agar plating of the dead bacterial samples confirmed complete absence of colony-forming units (i.e., viable bacteria) after IPA treatment.

**3.2. Preparation of proxy suspensions for sliding angle visualization:**

Proxy suspensions were prepared by doping either DI water or a DI water-ethylene glycol (EG)
binary mixture (mole fraction, $x_{EG} = 0.1$) with blue-colored polystyrene beads (PSB001UM,
MagSphere Inc.). To prepare the suspensions, 1 μL of the PSB001UM stock (10% w/v
concentration) was added to 1 mL of the corresponding base solution. The PSB001UM particles
are polystyrene latex beads dyed with an organic blue dye for enhanced visual detection. The
resulting suspensions had a particle concentration of approximately $10^7$ particles/mL. The average
particle diameter was 1 μm, with a diameter tolerance within 10% of the specified size.

**3.3. Transmission Electron Microscopy (TEM) Analysis:**

We used TEM to visualize and confirm the presence of flagella (responsible for motility) in the
live bacterial samples (see Figure S1 of the supporting information). 200 mesh F/C (Formvar +
Carbon) coated copper grids were used for sample preparation. The bacterial samples were applied
on the grids and left there for 5 minutes before the remaining solvent was removed using a
Whatman filter paper from below the grids. Then the grids were stained in 1% (wt/v) solution (20
µl) of ammonium molybdate for about 30 secs. After staining, the grids were left covered in the
fume hood to let them dry overnight before imaging. Transmission electron microscopy was
performed using a Philips CM10 microscope operated at an acceleration voltage of 60.0 kV.
Images were captured using an AMT 11-megapixel side-mounted high-resolution camera.

**3.4. Substrate preparation:**

In the present study, the wetting and adhesion experiments were conducted on superhydrophobic
substrates. These substrates were prepared by spray-coating clean glass slides with a two-part
multi-surface liquid repellent coating (NeverWet®, Rust Oleum). The base and top coats were
applied sequentially at ~ 30 min interval, followed by curing under ambient conditions for ~ 4 h.
Before coating, the glass slides were cleaned by ultrasonication in an acetone bath, followed by
rinsing with isopropanol and deionized (DI) water, and drying with compressed nitrogen to ensure
the base surface is clean before the coating is applied. Additional sliding (tilting plate) experiments
are conducted on PMMA surfaces, procured from Plaskolite Inc. Optically clear PMMA sheets
were supplied with polyethylene protective films on both faces. Sheets were stored in a closed
cabinet with the films intact to prevent contamination. Immediately before each experiment, the film was peeled away, and experiments were conducted on contamination-free pristine surfaces.
Solvent cleaning was avoided because many common solvents swell or damage PMMA.

**3.5. Substrate Characterization:**

The morphology and composition characterization of the superhydrophobic NeverWet samples
were carried out using Scanning Electron Microscopy (SEM) (see Figure S2 of supporting
information) and Energy Dispersive X-Ray (EDX) spectroscopy (see Figure S3 of the supporting
information) using a Quanta FEG 250 SEM with integrated EDS module at low-vacuum mode.

Additionally, we also performed stylus profilometry (P-6 stylus profiler, KLA Corp.) for both
PMMA and clean glass slides (scan length for line scan 1203 μm, force 1 mg). The surfaces were
sufficiently smooth, with glass slides registering an average roughness of 1.26 nm, while PMMA
recorded an average roughness of 40.16 nm. Therefore, it is safe to assume that the roughness has
minimal impact on the local pinning behavior.

**3.6. Environmental Conditions for Experiments:**

All the experiments were carried out at controlled laboratory conditions with an ambient
temperature of around $21 \pm 1$ °C and a relative humidity of around 60 %.

**3.7. Wetting & adhesion measurements:**

To measure the advancing and receding contact angles of live and dead bacterial suspensions on
the prepared substrate, we used a volume infusion/aspiration-based dynamic contact angle
goniometry (see top panel of Figure 1a) method using an optical goniometer (KRÜSS DSA 30,
USA). For advancing contact angle measurement, the test suspension was slowly infused onto the
substrate while the droplet profile was recorded using the side-view camera of the goniometer. The
advancing contact angle was defined as the contact angle measured at the advancing edge of the
droplet at the onset of outward motion of the three-phase contact line. For receding contact angle
measurement, liquid was gradually withdrawn from the droplet, and the receding contact angle
was defined as the contact angle measured at the receding edge at the onset of inward motion of
the three-phase contact line. Additionally, to further characterize contact-line mobility, we
performed tilting-plate measurements (see bottom panel of Figure 1a) on polymethyl methacrylate
(PMMA) substrates, which have relatively higher surface energy than the superhydrophobic
NeverWet®-coated glass surfaces. These measurements allowed both the dynamic

(advancing/receding) contact angles and droplet sliding behaviour to be characterized. In this method, a test droplet was placed on an initially horizontal platform, which was gradually tilted at a specified rate of $\sim 10°/\mathrm{min}$, and the contact-line behaviour of the droplet was recorded until the entire droplet slid down the inclined substrate.

For measuring the force of adhesion between the bacterial droplets and the test substrate, we used a cantilever-deflection-based adhesion characterization framework (see Figure 1b). A polymeric capillary tube of diameter 360 μm and spring constant, $k = 9.84\,\mu\mathrm{N/mm}$ is used as the cantilever. In the cantilever method, a probe droplet of the test liquid (here composed of live and dead *E. coli* suspensions, droplet volume $\approx 5\ \mu l$) is attached to the tip of the flexible cantilever, and then a series of sequential measurement steps is performed. Initially, the probe droplet-cantilever assembly is at its equilibrium undeflected position, not in contact with the target substrate. Once the probe droplet is dispensed at the tip of the cantilever, unwanted air-drag-induced perturbations of the cantilever are minimized, allowing reasonably accurate identification of this equilibrium position. Then, the target surface (here superhydrophobic NeverWet® coated glass surface, see subsection 3. 4 for details of surface preparation) is brought into contact with the probe droplet. As the target surface approaches the probe droplet, the cantilever-mounted probe droplet snaps onto the target surface (the "snap-in" stage). The magnitude of cantilever deflection during this stage depends on the affinity between the droplet and the test substrate, which is typically low for low-surface-energy superhydrophobic surfaces, as in the present study. Subsequently, the substrate is held in contact with the probe droplet for a predetermined duration (here ~ 10 *s*, the "holding" stage), after which it is retracted, leading to deflection of the cantilever during the "retraction" phase. Note that before the cantilever-mounted probe droplet detaches from the surface, its triple contact line gets depinned from the surface. Depending on the wettability of the surface and the droplet-substrate affinity, a measurable difference may exist between the cantilever deflection at the point of depinning and the eventual detachment of the droplet from the substrate. In that case, the deflection of the cantilever at the instance of depinning of the probe droplet should be considered for quantifying characteristic adhesion between the probe droplet and the test substrate, which, as we have shown in our previous work[50], coincides with the occurrence of zero acceleration of the cantilever tip during the retraction phase and can be determined via time-resolved motion analysis of cantilever tip. However, for the low wettability superhydrophobic surface used here, the detachment of the cantilever-mounted probe droplet from the test substrate almost coincides with the depinning of the probe droplet on the test substrate and therefore, the
experimentally observed point of maximum deflection, $\Delta X$, at the point of detachment can be used
directly in conjunction with the stiffness of the cantilever, $k$ to quantify the characteristic surface
adhesion, $\boldsymbol{F}_{adh}$ of the probe droplet as $\boldsymbol{F}_{adh} = k\Delta X$. The procedure for adhesion measurement is
discussed in greater detail in our previous works[48–51].

**3.8. Flow analysis:**

To explore the internal dynamics, the live/dead *E. coli*-laden droplets were monitored in a
confocal microscope (Zeiss LSM 800) for μ-PIV analysis (see Figure 1c). The droplet volume was
≈ 3 μl. The bacterial cells were stained with fluorescent stain for direct visualization. The staining
protocol involved mixing equal volumes of SYTO 9 dye and propidium iodide
(LIVE/DEAD *Bac*Light Bacterial Viability Kit, ThermoFisher, USA) in a micro-centrifuge tube.
Next, 3 μl of the dye mixture was added to each mL of bacterial suspension. The solution was
mixed thoroughly and incubated at room temperature for approximately 15 minutes in the absence
of light to prevent photobleaching. The μ-PIV experiments were conducted on clean glass slides.
The same protocol, as mentioned above in Substrate preparation (ultrasonication in acetone,
rinsing with isopropanol and DI water, and drying with compressed nitrogen), was followed for
cleaning the slides. Fluorescence from live and dead bacterial samples was collected through a
band-pass filter set, which selectively allowed signals within the desired wavelength range to pass
through while blocking unwanted signals (Figure 1(c)).

For μ-PIV image acquisition, the inter-frame time, $\delta_t$, was set to ~ 465 ms for live *E. coli*
suspensions, and ~ 233 ms for dead *E. coli* suspensions, corresponding to effective acquisition
rates of approximately 2.15 and 4.30 frames per second, respectively. The shorter inter-frame time
for dead *E. coli* suspensions was selected because these dead cells were seen to be more strongly
advected by evaporation-driven capillary flow and therefore exhibited higher apparent flow
speeds. This choice ensured the inter-frame displacement remains within a range suitable for cross-
correlation. The local velocity field of the bacterial suspension was measured using a cross-
correlation-based PIV algorithm (PIVLab, MATLAB). For the concentration used, there was a
significant overlap of cells, so it was not possible to resolve individual bacterial velocities. For
PIV analysis, each successive image pair was divided into overlapping interrogation windows,

each containing multiple bacteria/cells. The displacement within each interrogation window was estimated by cross-correlation with the corresponding region in the successive image, yielding a two-dimensional velocity field. In the present study, each interrogation window size was 32 pixels x 32 pixels, corresponding to approximately 19.94 μm x 19.94 μm, with 50% overlap between neighbouring windows. This produced a 31 x 31 vector field over the analyzed observation window of 512 pixels x 512 pixels (corresponding physical dimensions ~319 μm x 319 μm). All images were pre-processed, where the background intensity was subtracted to increase contrast. A uniform 3 × 3 smoothing kernel was applied to reduce noise in the resulting velocity field. Also, the correlation was done with a time step of $\delta_t$ (inter-frame time spacing during image acquisition).

Note that the imaging depth of field[52], $\delta z$ can be estimated as:

$$\delta z = \frac{n_r \lambda_l}{NA^2} + \frac{n_r e}{M(NA)} \quad \text{Eq. (1)}$$

where $n_r \approx 1$ is the refractive index of the air between the sample and the objective, $\lambda_l \approx 0.488\,\mu m$ is the illumination wavelength, $e \approx 0.6\,\mu m$ is the minimum distance that can be resolved by the detector, and $M \approx 10$ is the magnification, and $NA = 0.3$ is the numerical aperture of the objective (EC-Plan Neofluar, Carl Zeiss, Canada). This gives $\delta z \approx 5.6$ μm. The characteristic velocity scale of the stained *E. coli* cells in our experiments is of the order of 1 μm/s. Therefore, over the selected inter-frame intervals ($\delta_t$), the expected bacterial displacement remains smaller than both the interrogation-window length scale and the estimated imaging depth of field, allowing reliable cross-correlation for μ-PIV analysis.

## 4. RESULTS AND DISCUSSION:

### 4.1. Research Concept and Outline of the Work:

The primary objective of this work is to understand the true nature of early-stage (~*O*(100s)) surface interaction of *E. coli*-laden droplets. All reported measurements were conducted within the first $\approx 100\ s$ after droplet-surface contact to ensure that the volume loss due to evaporation is not significant, and to capture the active motility phase of the bacterial suspension. This time frame was selected to isolate early-stage interfacial dynamics without interference from longer-term effects. A dual-scale approach was adopted (see Figure 1 for a schematic representation of the research concept) to achieve a holistic understanding of the ensuing
interfacial dynamics. First, the macroscopic surface interaction was investigated at a droplet scale.
Two different approaches are used to study the macroscopic interaction. Initially, the dynamic
wetting behaviour was examined via contact angle goniometry, where the advancing and receding
contact angles of *E. coli*-laden droplets were measured on a superhydrophobic surface (see the top
panel in Figure 1(a)). However, no conclusive/statistically significant difference is observed in the
dynamic wetting behaviour between live and dead *E. coli*-laden droplets. For brevity, the terms
"live *E. coli*-laden droplets" and "live droplets" (as well as "dead *E. coli*-laden droplets" and "dead
droplets") are used interchangeably throughout this manuscript.

The inconclusive nature of contact angle goniometry prompted us to switch to an alternate,
more accurate method for direct quantification of droplet-scale surface interaction: the cantilever-
deflection-based adhesion quantification method (Figure 1(b)), which can measure the force of
adhesion between a droplet of a test liquid and a target surface with micro-Newton (μN) accuracy
(see subsection 3.7 in Materials and Methods for the details of the cantilever method for adhesion
quantification). Interestingly, despite the inconclusive nature of the dynamic wetting signature
captured via contact angle goniometry, the cantilever method revealed a distinctive difference
between live and dead droplets. When a live *E. coli*-laden droplet was used as the probe droplet,
the maximum deflection of the cantilever was significantly lower compared to its dead counterpart
(see the bottom panel in Figure 1(b)), indicating distinctly lower surface adhesion of live droplets
compared to dead droplets. Note that the individual live and dead bacteria are similar in size, as
confirmed via dynamic light scattering (see Figure S4 of the supporting information). This
apparent anomaly prompted us to further investigate whether the internal motion dynamics of
suspended *E. coli* cells (live vs dead) could explain the measured adhesion contrast.

At the microscopic scale, the motion of fluorescently stained *E. coli* cells within DI water
sessile droplets was visualized (details in Materials and Methods) using micro-particle image
velocimetry (μ-PIV) (Figure 1(c)). Similar to the cantilever method, μ-PIV also captured a
prominent difference between live and dead *E. coli*-laden droplets, albeit at a different observation
scale. These observations reveal a marked difference in the internal motion dynamics in live and
dead droplets, suggesting that bacterial motility can significantly alter the motion of suspended
cells within the droplet and may contribute to the reduced surface adhesion observed in live
droplets.

To further test our hypothesis regarding the manifestation of internal motion dynamics on
bulk droplet-scale surface interactions, we performed additional wetting measurements, this time
using a substrate with relatively higher surface energy (PMMA). This was done with the
anticipation that the enhanced wettability of the PMMA substrate might allow distinctions in the
wetting signature between live and dead droplets to be captured. In this case, droplet sliding
measurements were performed using the tilting plate method (see subsection 3.7, Materials and
Methods). The sliding measurements confirmed enhanced contact line mobility for live droplets,
indicating that the motility-dependent internal motion observed by μ − PIV is accompanied by a
measurable change in droplet-scale contact-line dynamics. To the best of our knowledge, this is
the first report of differential contact-line mobility between live and dead *E. coli*-laden droplets.

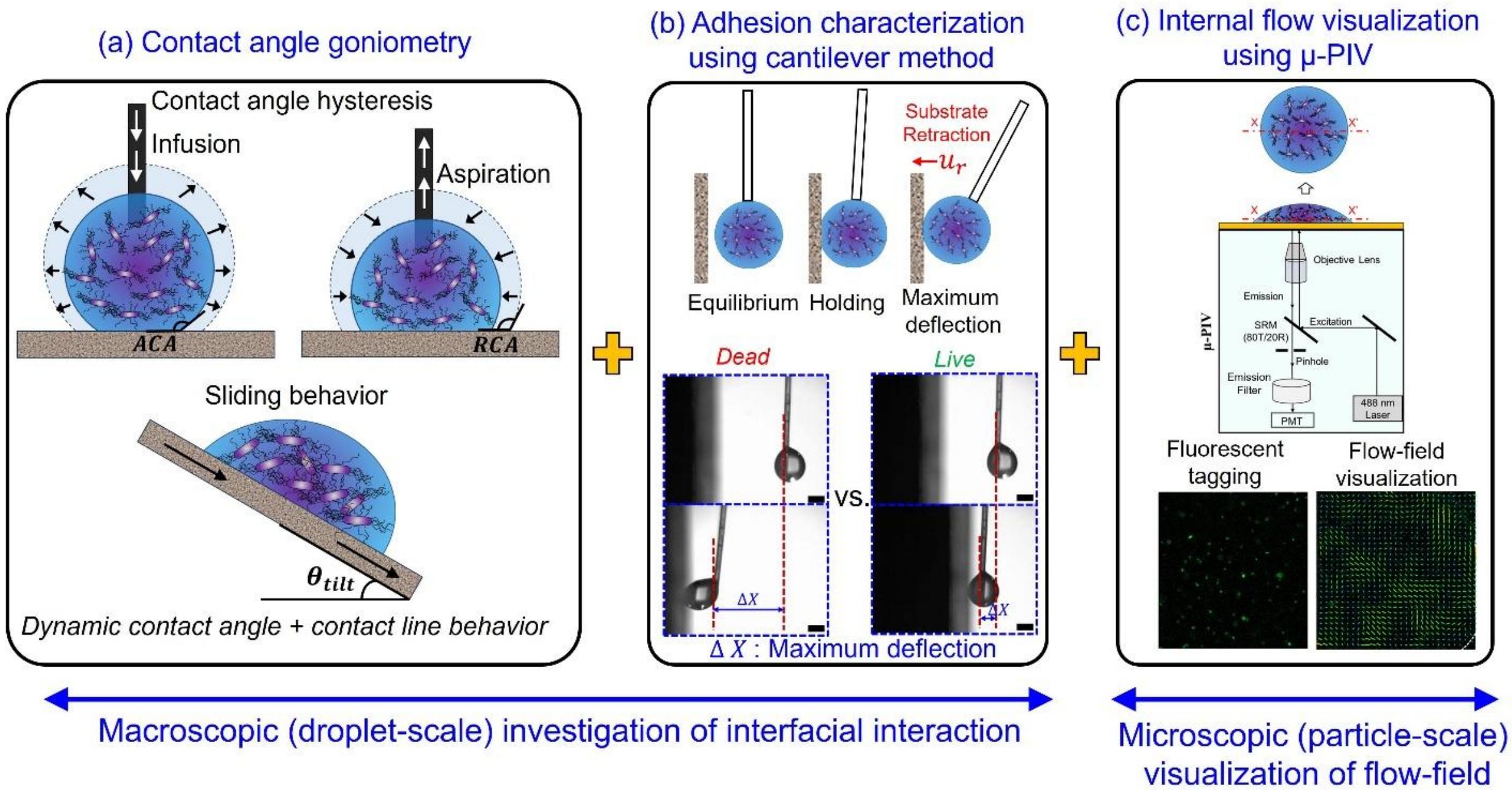

**Figure 1: Research Concept: A Dual-scale approach to understand the early-stage (~*O* (100
s)) surface interaction of *E. coli*-laden droplets.** Observations are conducted at two different
length-scales to decipher the true nature of the surface interaction. (a-b) Macroscopic (droplet-
scale) investigation comprises (a) contact angle goniometry (via volume infusion/aspiration or
tilting plate method) to characterize the dynamic contact angle and the contact line behavior and
(b) direct quantification of surface adhesion force using a cantilever-based method. (c) At
microscopic (particle) scale, the investigation consists of fluorescent staining of bacteria and
subsequent micro-particle velocimetry (μ-PIV) observation to visualize the internal flow fields of
live/dead *E. coli*-laden droplets.

### 4.2. Macroscopic (droplet-scale) assessment of surface interaction of bacterial droplets

As described in the previous section, the macroscopic evaluation of surface interactions of *E. coli*-laden droplets is performed using two different approaches: dynamic wetting characterization via contact angle goniometry and direct quantification of surface adhesion using a cantilever deflection method. In addition to comparing the surface interactions of live and dead droplets, the effect of bacterial cell concentration is also investigated. The undiluted bacterial stock has an estimated concentration $\approx 10^7$ CFU/ml. For both live and dead cases, two additional concentrations are prepared through 10-fold serial dilutions. The surface interactions of all three concentrations are studied to achieve a holistic understanding of how both the concentration and the nature (live vs. dead) of the suspended particles influence interaction behavior.

#### 4.2.1. Dynamic Wetting Characterization via Contact Angle Goniometry:

Before performing dynamic wetting characterization, we first measured the surface tension values of live vs dead droplets. As shown in Figure 2 (a), the surface tension of dead droplets closely resembles that of the bulk suspension media, i.e., deionized (DI) water, with no statistically significant variation in surface tension as the concentration of the bacterial cells is increased in the dead droplets. In contrast, the live droplets exhibited relatively lower surface tension, with the deviation from the suspension medium becoming more prominent at higher bacterial concentrations. At the lowest tested concentration (~ $10^5$ CFU/ml), the surface tension (72.37±0.06 mN/m) was almost identical to DI water, reducing slightly (71.94±0.1 mN/m) with a ten-fold increase in bacterial concentration to ~ $10^6$ CFU/ml, which was statistically insignificant. However, at ~ $10^7$ CFU/ml, the surface tension significantly decreased to 65.82±1.6 mN/m. This outcome is consistent with the fact that live *E.coli* bacteria is capable of producing biosurfactant.

Next, we evaluated the dynamic wetting behavior of the *E. coli*-laden droplets on a superhydrophobic (NeverWet®) surface, which records an advancing contact angle (ACA) ~ 163.4±2.8° and receding contact angle (RCA) ~ 158.2±3° for deionized (DI) water. The volume infusion/aspiration method (see the schematic in Figure 1 (a) for a visual representation of the process and the methods section for the experimental details) is used for dynamic wetting characterization. No prominent distinction was found in the dynamic wetting signature between live and dead droplets. Live droplets recorded (see Figure 2 (b)) ACA values of 168.2±2.9°,

168.5±2.5°, and 168.4±3.9°, and RCA values 155.7±3.8°, 160.6±4.3°, and 153±7.8°, for the three
tested concentrations (~$10^5$, $10^6$, and $10^7$ CFU/ml, respectively). Dead droplets exhibited (see
Figure 2 (c)) ACA values 167.7±1.8°, 169.5±2.1°, and 167.2±2.8°, and RCA values 151.2±4.1°,
154.5±3.8°, and 151.6±4.6° for the same three tested concentrations, in the above-mentioned order.
Statistical comparisons across concentrations were performed separately for ACA, RCA, and CAH
using one-way ANOVA followed by Tukey-Kramer post hoc tests. Neither live nor dead *E. coli*-
laden droplets exhibited any statistically significant concentration-dependent trend across the
tested concentrations.Demonstrative experimental snapshots of the advancing and the receding
stage during dynamic wetting characterization of both live (Figure 2(d)) and dead (Figure 2(e))
droplets are also presented for all three tested concentrations, which confirm that contact angle
goniometry does not provide a robust basis for differentiating between live and dead *E. coli*-laden
droplets or their varying concentrations in terms of their droplet-surface interaction on
superhydrophobic NeverWet® surfaces. Although this result might seem surprising given the
lower surface tension of live *E. coli*-laden droplets compared to their dead counterparts (Figure
2(a)), the inconclusive nature of the dynamic wetting signature is not entirely unexpected, given
the fact that the measurements were performed on a superhydrophobic surface. Image-based
contact angle goniometry on superhydrophobic surfaces suffers from significant imaging
challenges arising from optical noise caused by scattering and diffraction near the three-phase
contact line. It can lead to inaccurate identification of the baseline of the droplet, thereby causing
the propagation of substantial systematic errors[53] in measurement. It is worth mentioning that on
superhydrophobic surfaces, an error of one pixel in baseline identification can lead to almost a 300
% error[53] in estimating the adhesion force.

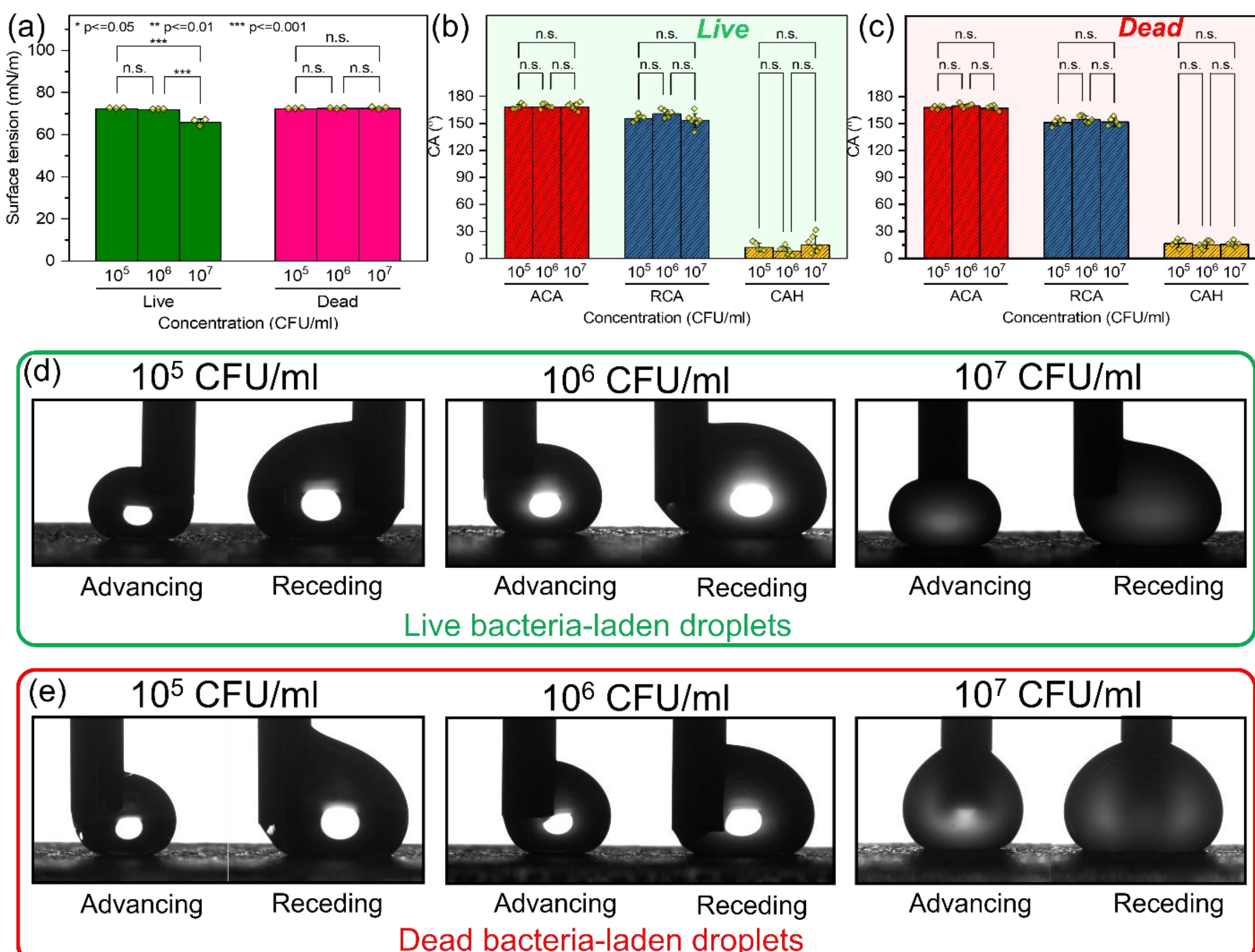

**Figure 2: Dynamic Wetting Characterization of *E. coli*-laden droplets:** (a) Surface tension values for live and dead *E. coli*-laden droplets across the three tested concentrations. (b, c) Advancing contact angle (ACA), receding contact angle (RCA), and contact angle hysteresis (CAH) values recorded on the superhydrophobic NeverWet® surface for live (b) and dead (c) droplets at the three tested concentrations. (d, e) Experimental snapshots of the advancing and receding stages during dynamic contact angle measurements for live (d) and dead (e) droplets across the three tested concentrations. For all the reported column plots, the height of each column represents the corresponding mean value, with error bars indicating ± one standard deviation. Individual raw data points are overlaid to show sample-level variability. Statistical comparison across concentrations were performed using one-way ANOVA followed by Tukey-Kramer post hoc tests, with corresponding significance levels indicated within the plots. n.s.: not significant, $* \, p \leq 0.05$, $** \, p \leq 0.01$, and $*** \, p \leq 0.001$.

#### 4.2.2. Direct quantification of surface adhesion using cantilever method:

The limited ability of contact angle goniometry to distinguish between live and dead *E. coli*-laden droplets prompted us to adopt a direct and more accurate, cantilever-based method[47–51] for the quantification of droplet-surface adhesion. The cantilever-deflection method for adhesion quantification is discussed in Materials and Methods and also in our previous works[48–51].

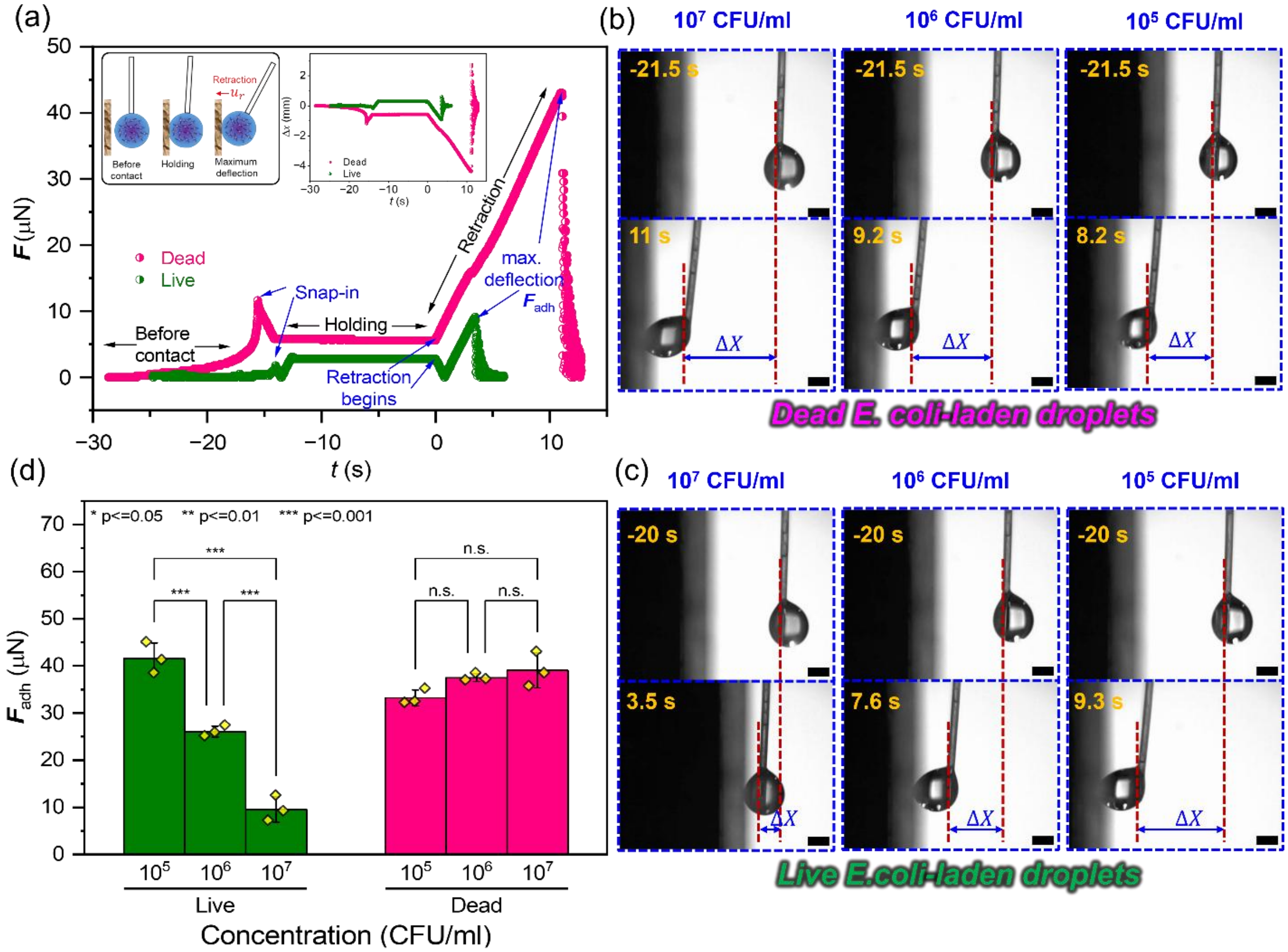

**Figure 3: Direct Quantification of Surface Adhesion Using Cantilever Deflection Method**: (a) Typical force (F) vs. time (t) curves for live and dead *E. coli*-laden droplets with a concentration of $\approx 10^7$ CFU/ml, presented side-by-side, during a typical surface adhesion measurement sequence. The left inset schematically illustrates the stages of the measurement sequence, while the right inset shows the corresponding displacement ($\Delta x$) vs. time (t) curves. (b, c) Experimental snapshots of the equilibrium and maximally deflected states of the cantilever during adhesion measurements for dead droplets (b) and live droplets (c) at all three tested concentrations. The top panels display the equilibrium state, while the bottom panels show the maximally deflected state, highlighting the maximum deflection ($\Delta X$). Scale bar in the snapshots presented in (b, c) is 1 mm. (d) Grouped column plot summarizing the adhesion data for all three concentrations and both live and dead droplets. The height of each column represents the corresponding mean value, with error bars indicating ± one standard deviation. Individual raw data points are overlaid to show sample-level variability. Statistical relevance was analyzed using one-way ANOVA followed by Tukey-Kramer posthoc test, with corresponding significance (p-value) levels indicated within the plot.

Figure 3(a) shows representative force-time curves obtained during adhesion measurements for live and dead droplets at an estimated concentration of $\sim 10^7$ CFU/mL. The left inset schematically presents the different stages of the measurement, and these stages are marked on the force-time curves as well. The right inset shows the corresponding cantilever displacement vs.
time curves. The droplet-surface adhesion force, $\boldsymbol{F}_{adh}$, was obtained from the maximum cantilever
deflection, $\Delta X$, during substrate retraction as $\boldsymbol{F}_{adh} = k\Delta X$, where $k$ is the stiffness of the
cantilever. Figures 3(b,c) present experimental snapshots of the equilibrium and maximum-
deflection states for dead and live *E. coli*-laden droplets, respectively, across all tested
concentrations. For each concentration, the top panel shows the equilibrium position, while the
bottom panel marks the maximum-deflection state, with the separation between these two positions
giving $\Delta X$. These snapshots reveal an apparent dependence of $\Delta X$ on concentration, the nature of
which varies depending on whether the suspended *E. coli* cells are live or dead. The concentration
dependence of $\boldsymbol{F}_{adh}$ for both live and dead droplets is summarized in Figure 3(d) using grouped
column plots. Dead droplets show a slight increment in $\boldsymbol{F}_{adh}$ with increasing concentration, which
is found to be statistically not significant. In contrast, the live droplets demonstrate a statistically
significant decrease in $\boldsymbol{F}_{adh}$ with increasing concentration (p value $\leq 0.001$). Comparative
experimental videos showing the concentration dependence of surface adhesion for live and dead
*E. coli*-laden droplets are provided in Supporting Videos S1 and S2, respectively.
Occurrence of a definitive trend in $\boldsymbol{F}_{adh}$ might appear counterintuitive, given the
inconclusive results from contact angle goniometry on the same superhydrophobic substrates.
However, such outcomes are not uncommon. For example, a theoretical estimation of work of
adhesion using the Young-Dupré equation highlights the inherent limitations of contact angle
measurements on superhydrophobic surfaces. Using a typical set of experimental parameters
relevant for live droplets with a concentration $\approx 10^7$ CFU/ml (surface tension $\gamma \approx 65\ \frac{\mathrm{mN}}{\mathrm{m}}$ and
$RCA \approx 150°$), the calculated work of adhesion is: $W \approx \gamma(1 + \cos RCA) \approx 8.71\ \frac{\mathrm{mJ}}{\mathrm{m}^2}$. However,
an underprediction of RCA by just 5°(i.e., $RCA \approx 145°$), an error typical[53] for superhydrophobic
surfaces, leads to $W \approx 11.75\ \frac{\mathrm{mJ}}{\mathrm{m}^2}$, resulting in an error of ~35%. The same argument applies to the
force of adhesion, $\boldsymbol{F}_{adh}$ as well. This highlights the severity of systematic errors in contact angle
measurements on superhydrophobic surfaces and points out the unsuitability of contact angle
goniometry as an indirect tool to quantify droplet-surface interactions, especially when subtle
variations in the interaction forces, often µN range, are involved. In contrast, the cantilever method
is significantly more sensitive towards such subtle yet definitive variations and offers the
possibility of improving the sensitivity via tailoring the stiffness of the cantilever. This explains why the cantilever method successfully detected a trend in adhesion signature, while contact angle method turned out to be inconclusive.

The slight concentration-dependent increase in surface adhesion for dead droplets can be attributed to the inert nature of dead bacterial cells, which act as pinning barriers at the contact line. With increasing concentration, this leads to enhanced contact line pinning. It restricts the mobility of the contact line, and lead to an increased $\boldsymbol{F}_{adh}$. Similar pinning effects with inert microparticle-laden suspensions have been reported previously[49,54].

A more intriguing observation is the noticeable decrease in $\boldsymbol{F}_{adh}$ for live droplets with increasing bacterial concentration, despite the reduced surface tension of live bacterial suspensions at higher concentrations. From a conventional standpoint, lower surface tension of the probe droplets would be expected to result in higher adhesion forces. This contradiction raises some critical questions: *Why does the presence of live bacteria reduce surface adhesion? Why does adhesion further decrease with increasing bacterial concentration? What internal droplet dynamics are at play in live droplets to cause this behavior?*

These questions are explored in the subsequent sections, where we examine the internal motion dynamics of suspended bacterial cells at a microscopic (particle) scale within sessile droplets and investigate how bacterial migration behavior influences the observed trends.

### 4.3. Microscopic visualization of internal motion dynamics of *E. coli*-laden droplets:

Given the counterintuitive trend of surface adhesion observed in Figure 3, we attempted to visualize the internal flow field of bacterial cells to determine its potential influence on surface interactions. The anomalies in the surface adhesion of live droplets may stem from the migration behavior of the suspended *E. coli* bacteria within. Therefore, to understand surface interactions and identify key differences in wetting behavior between live and dead droplets, it is crucial to investigate the flow field of the self-propelled, motile bacteria suspended within the droplets. In the present study, we use $\mu-\text{PIV}$ to probe the flow field of the suspended bacterial cells within the sessile droplets. The $\mu-\text{PIV}$ of the sessile droplets were performed on transparent glass substrates to enable bottom-view visualization. Details of staining methodology and the $\mu-\text{PIV}$ protocol are provided in the methods section. Micro-particle image velocimetry measurements were performed for bacterial concentrations of $10^7\ \text{CFU/ml}$ and $10^6\text{CFU/ml}$ only. Measurements for $10^5$ CFU/ml were not feasible due to the low particle concentration within the sessile droplet. Flow fields were observed near the contact line, approximately 15 μm from the substrate wall, with an observation window of 319 μm × 319 μm. See Supporting Videos S3 and S4 for the time-resolved evolution of the internal velocity fields of live and dead bacteria within sessile droplets, respectively, at a bacterial concentration of $10^7$ CFU/ml. The μ-PIV analysis is performed in a Cartesian framework. The velocity field, therefore, has two components, namely, the $U$ and $V$ components, aligned along the $x$ and $y$ axes of the frame of reference, respectively. Owing to the radial symmetry of the spherical cap-shaped sessile droplet, the velocity statistics for both the $U$ and $V$ components are expected to remain qualitatively similar.

Figure 4(a) presents the time-resolved $x$-component of velocity ($U$-velocity) of bacterial cells as a surface plot, while Figure 4(b, c) shows illustrative snapshots of the overall velocity vector fields and streamline patterns for live and dead droplets at a concentration of $10^7$ CFU/ml. Positive (+) $U$-velocity values represent motion toward the triple contact line (TCL), while negative (-) values indicate motion away from TCL. We will come to the directionality of the velocity vectors in detail in the subsequent section. However, at this point, we would like to draw the attention of the readers to the $Z$-values of the surface plots presented in Figure 4(a). It highlights a clear trend: in general, dead bacteria exhibit higher positive $U$-velocities (i.e., toward TCL velocity) compared to their live counterparts. For direct visual comparison of velocity magnitudes between live and dead droplets at matching timepoints, the corresponding surface plot pairs are presented on common axes in Supporting Information Figure S5. The comparative velocity vector fields and the streamline patterns presented in Figure 4(b, c) reveal the same fundamental difference in motion characteristics between live versus dead bacteria. Dead bacteria exhibit consistent motion toward the contact line, resembling the behavior of inert particles. This motion is likely driven by capillary flow generated by evaporative flux within the droplet. In contrast, live bacteria do not have a predominant tendency to move toward the contact line. Under the same experimental conditions, the flow velocities of live bacteria towards the TCL are noticeably smaller, suggesting that live bacteria resist being passively driven to the contact line by evaporation-induced capillary flow.

Note that the flow velocity obtained from $\mu - \text{PIV}$ is the resultant flow-field demonstrated by the bacterial cells. The net resultant flow velocity for live droplets, $\vec{v}_{net,l}$ can be expressed as a vector sum of the flow-field of the capillary flow, $\vec{v}_{cap}$ and the internal (flagellar motility-driven) flow field of live bacteria, $\vec{v}_{i,l}$, i.e., $\vec{v}_{net,l} = \vec{v}_{cap} + \vec{v}_{i,l}$.

Given that the dead bacteria are incapable of generating their own flow field, the net flow field observed in dead droplets can be considered representative of the capillary flow field. The observation that the net flow field of live droplets is consistently lower in magnitude than that of dead droplets suggests that live bacteria actively generate their own internal flow field, which opposes the evaporation-driven radially outward capillary flow (i.e., directed towards the TCL) inside the sessile droplet placed on a hydrophilic glass surface. Quantitative measurements supporting these observations are presented in the subsequent section.

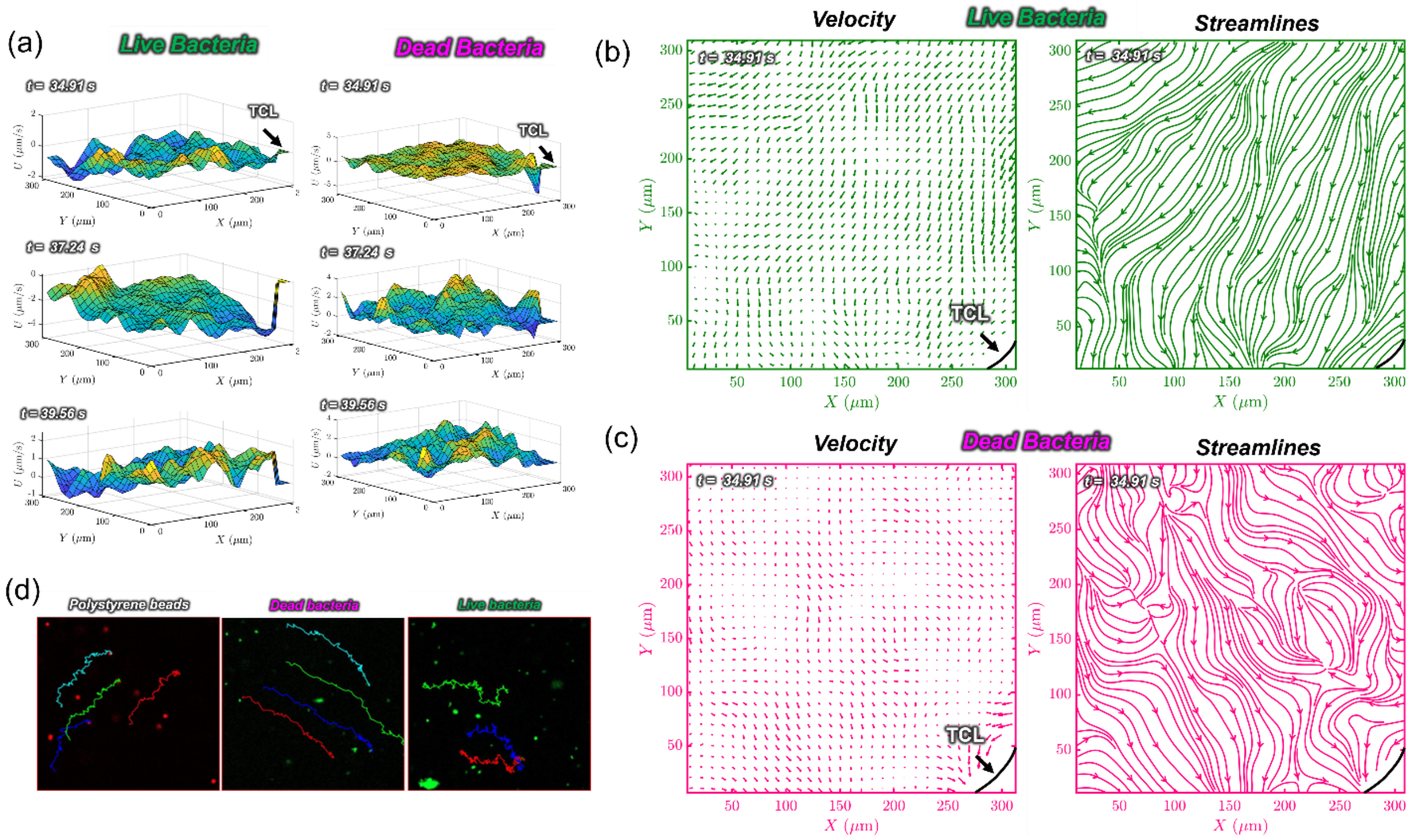

**Figure 4: Visualization of the internal flow-field of *E. coli*-laden droplets using micro-particle image velocimetry (μ-PIV)** - (a) shows the spatial distribution of the $x$-component of the velocity vector (i.e., the $U$ velocity) in a time-resolved manner for both live and dead droplets. TCL indicates the triple contact line of the sessile droplet. (b-c) show illustrative velocity vector field and the streamline patterns for both (b) live and (c) dead droplets. (d) displays the constructed pathlines of suspended particles within a sessile DI water droplet for three cases: polystyrene beads, dead bacteria, and live bacteria, shown side by side. Each trajectory in (d) is obtained by manually tracking individual tracers over time. For the live case, 200 frames were tracked with an interframe spacing of ≈ 930.91 ms (total duration ≈ 186.18 s); for dead bacteria, 100 frames

with the same interframe spacing were used ($\approx$ 93.09 s); for polystyrene beads, 100 frames were processed at $\approx$ 297.89 ms spacing ($\approx$ 29.78 s total duration). In each panel, a minimum of three individual tracers were followed. Longer track durations were feasible for live bacteria due to their lower net velocity and fluctuating motion. The bacterial concentration for all the cases represented in Figure 4 is $\approx 10^7$ CFU/ml. The time labels shown in panels (a-c) coincide exactly with the corresponding timestamps in Supporting Videos S3 (live) and S4 (dead), enabling direct cross-referencing.

To further confirm that live bacteria do not conform to induced capillary flow, we conducted an additional set of experiments comparing the motion dynamics of inert fluorescent polystyrene beads, dead *E. coli* cells, and live *E. coli* cells suspended within sessile DI water droplets on glass substrates. Individual trajectories were tracked over time using Lagrangian particle tracking, and representative pathlines were constructed to present the motion of individual particles/cells (see Figure 4 (d)). As expected, the evaporation-driven capillary flow in sessile DI water droplets tends to drive all suspended inert particles, including fluorescent polystyrene beads and dead bacteria, radially outward toward the contact line. However, the propensity of live *E. coli* bacteria to move toward the contact line is considerably lower. Live bacteria exhibit unique and somewhat haphazard motion patterns: a continuous and prominent change in directionality of motion could be observed from the constructed pathlines, and crossover between the pathlines of two neighbouring bacteria could also be observed, suggesting interactions among individual bacteria. Their trajectories also show a tendency to resist being carried downstream, indicative of positive rheotaxis. These interactions and their implications are further explored and discussed in the subsequent section.

### 4.4. Statistical analysis of the motion dynamics:

In the preceding discussion, we have outlined the flow behavior of the bacterial cells acquired via μ-PIV measurements. In doing so, we provide an Eulerian description of the velocity field (i.e., description of the velocity statistics at defined grid points) over a reasonably large observation window near the triple contact line (TCL) of the sessile droplet. In parallel, we performed individual particle tracking to obtain a complementary Lagrangian perspective, which provided hints of some interactions occurring among the motile *E. coli* bacterial population. Together, these observations motivated a more systematic statistical quantification of the spatiotemporal dynamics observed in both live and dead bacterial suspensions. The analysis that follows integrates both vector components of the velocity field to probe directional preferences, net transport tendencies, and the effect of bacterial motility on coordinated flow behaviour.

#### 4.4.1. Directional Trend of the Velocity Field

To understand the cumulative motion dynamics, we first look at the time-resolved non-normalized histograms of the $U$ −velocities of the bacterial cells (See Figure 5(a)-(b)). An estimated $\approx 10^7$ CFU/ml concentration is maintained for both live and dead samples. Figures 5 (a), (b) show that the histograms can be reasonably approximated using a normal (Gaussian) distribution. However, a distinctive trend could be detected by comparing the histogram profiles between the live and dead bacterial suspensions. For instance, in dead samples, most of the velocity vectors have a positive magnitude, indicating a motion predominantly towards the contact line, which conforms with the evaporation-driven capillary flow, as shown in Figure 4 and discussed in the previous subsection. On the contrary, in live samples, the distribution of $U$ velocity shifts around $U = 0$ with time. It indicates that the percentage distribution of the number of bacterial cells that have a resultant motion towards the TCL vs. the ones directed away from the TCL continuously evolves, leading to the observed haphazardness in the motion, as already discussed in Figure 4 (d).

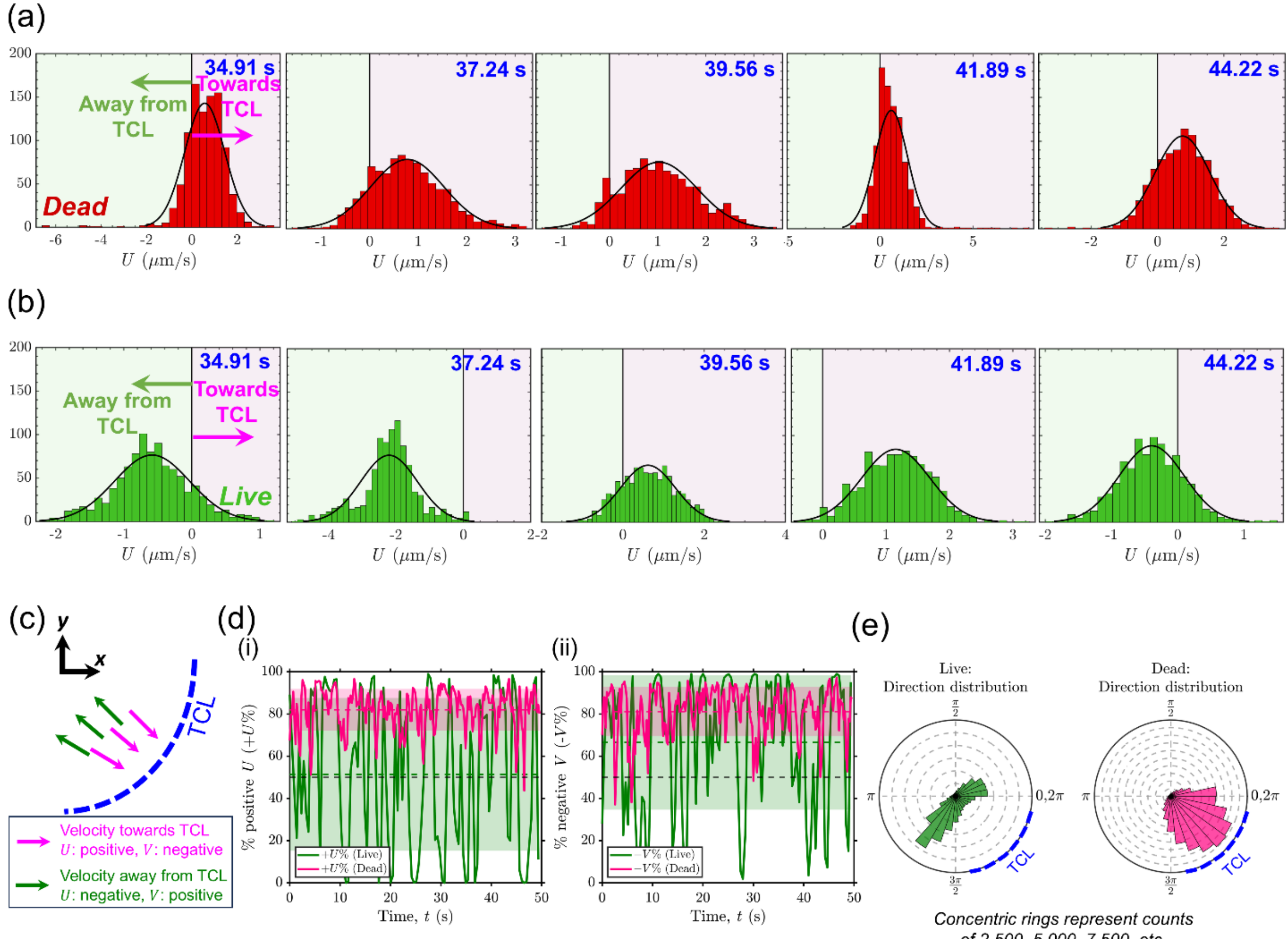

**Figure 5: Directional statistics of bacterial motion inside a sessile droplet**- (a), (b) present time-resolved $U$-velocity histograms for sessile droplets containing the dead and live bacterial loads, respectively. The time points correspond to uniformly spaced intervals of approximately 2.33 s for both cases. The time labels shown in panels (a-b) coincide exactly with the corresponding timestamps in Supporting Videos S4 (dead) and S3 (live), enabling direct cross-referencing. (c) Schematic of the coordinate system used to define directionality of the velocity components. (d) Temporal evolution of directional preference- (i) Time-resolved percentage of velocity vectors with positive $x$ components ($+U\%$), and (ii) percentage of vectors with negative $y$ components ($-V\%$), both indicating motion toward the TCL along their respective axes. Each curve reflects the temporal behavior over a $50\,s$ observation window. Solid green/magenta curves show the instantaneous $+U\%$ (i) and $-V\%$ (ii) for live/dead cases. The colored dashed lines mark the time-averaged values $\langle +U\% \rangle$ and $\langle -V\% \rangle$, and the semi-transparent bands span ±1 temporal standard deviation around those means. The black dashed horizontal line at the $50\%$ mark serves as a reference line for direction reversal. Live samples show strong temporal fluctuations around 50%, whereas dead samples exhibit consistently high values for both metrics, indicating stable capillary-driven flow. (e) Polar histogram showing the angular distribution of all velocity vectors over the entire spatiotemporal domain. Radial direction indicates frequency; angular bins are $\frac{\pi}{15}$ wide. The fourth quadrant ($\frac{3\pi}{2} - 2\pi$), representing vectors with $U > 0$ and $V < 0$ (i.e., directed toward the

TCL), is prominently populated in dead samples but is notably suppressed in live ones. The
concentration is kept at ~ $10^7$ CFU/ml for both live and dead bacteria.

To precisely quantify the intrinsic directional dynamics observed in the bacterial droplets,
it is essential to clearly define the reference coordinate system, as we do in the schematic (Figure
5(c)). Our coordinate axes are aligned such that velocity vectors with both a positive $x$ component
($+U$) and a negative $y$ component ($-V$) represent motion clearly directed towards the triple contact
line (TCL), with opposite directions ($-U$, $+V$) corresponding to movement away from the TCL.
Using this coordinate convention, we further quantify the directional bias of the velocity
field in live vs dead bacterial samples. We compute the percentage of velocity vectors with positive
$x$ component ($+U\%$) and negative $y$ component ($-V\%$) at each time frame over the full
observation window of 50 s. These percentage values are presented as time series in Figure 5(d)(i-
ii), allowing us to examine the temporal evolution of directional preference toward the TCL at the
level of individual components of the overall velocity vector field. Specifically, Figure 5(d)(i)
shows the temporal evolution of the proportion of velocity vectors with positive horizontal
component ($U > 0$), while Figure 5(d)(ii) presents the corresponding percentage with negative
vertical component ($V < 0$). For dead bacterial suspensions, both metrics remain consistently high
throughout, with a time-averaged $+U\%$ of $81.96 \pm 9.89\%$ and $-V\%$ of $81.11 \pm 11.65\%$,
indicating persistent motion toward the TCL driven by evaporation-induced capillary flow. In
contrast, the live bacterial samples exhibit strong temporal fluctuations, with time-averaged values
of $51.37 \pm 36.18\%$ for $+U\%$ and $66.54 \pm 31.88\%$ for $-V\%$. These fluctuations are centered
near the $50\%$ reference line, with frequent crossovers, highlighting the dynamic competition
between active bacterial motility and the imposed capillary flow. The net effect of this competition
is frequent reversals in the directional orientation of the velocity field of the live bacterial
population. The temporal changes in the directionality of $U$-velocity discussed here can also be
visualized in Supporting Videos S3 (live) and S4 (dead), where the right panel of each frame color-
codes the sign of $U$-velocity at each PIV grid point. Red denotes motion toward the contact line
($U > 0$), blue indicates motion away ($U < 0$), and black regions demarcate areas outside the
droplet. This frame-by-frame visualization enables direct spatiotemporal comparison of the
evolving directional trends.
While Figure 5(d) tracks the temporal evolution of directional bias by reporting the
framewise percentage of vectors with positive $U$ or negative $V$ components, one component at a time, we attempt to provide a more direct and comprehensive measure of directionality by directly
analyzing the angular distribution of all instantaneous velocity vectors over the entire
spatiotemporal ensemble in Figure 5 (e). The polar histogram shown in Figure 5(e) presents the
frequency of velocity directions, where the orientation of each vector is computed based on its
instantaneous $(U, V)$ components. The angular range spans $0 - 2\pi$ in a counterclockwise manner,
with each bin width representing an angular span of $\frac{\pi}{15}$. We are particularly interested in the fourth
quadrant ($\frac{3\pi}{2} - 2\pi$), which exclusively corresponds to velocity vectors that simultaneously satisfy
$U > 0$ and $V < 0$, i.e., velocity vectors those are clearly directed toward the three-phase contact
line (TCL) (see sign convention in Figure 5(c)). To quantify this directional bias, we compute the
total number of such vectors across the entire domain. In the data presented in Figure 5 (e), for
dead bacterial suspensions, $139{,}285$ out of $205{,}654$ vectors ($31 \times 31$ spatial grid, total $214$
timeframes, inter-frame spacing, $\Delta t = 0.23273\ s$, observation window $\approx 49.80\ s$) satisfy both the
$U > 0$ and $V < 0$ condition, yielding a net proportion of 67.73% that lies in the fourth quadrant.
In contrast, in live bacterial suspensions only $24{,}305$ out of $102{,}827$ total vectors (same grid size,
$107$ frames, $\Delta t = 0.46545\ s$, resulting in the same observation window of $\approx 49.80\ s$ as dead
samples) lie in the fourth quadrant, corresponding to just 23.64% velocity vectors being directed
towards TCL. Importantly, the angular distribution for live samples is not only lower in magnitude
within the TCL-directed quadrant but also substantially broader in spread, indicating frequent
directional reorientations. This behaviour, when viewed alongside the temporal directional
instability observed in Figures 5(d)(i-ii), reveals that motile bacteria do not passively follow
capillary-driven flow but instead dynamically reorient to modulate or oppose it.

Together, the component-wise variations in Figure 5(d) and the quadrant-specific
preferential directional alignment patterns in Figure 5(e) present a coherent picture: while motile
bacteria may transiently align with the TCL-directed flow, their collective dynamics frequently
disrupt this directionality, reflecting an active resistance to capillary transport driven by their
motility.

**4.4.2. Net Velocity Trends and Temporal Fluctuations**

While the previous analysis (Figure 5) focused on the directionality of the velocity field, it
is equally important to assess the magnitude and net directional bias of the flow. For this, we compute the spatially averaged signed velocity components at each time frame over the entire
observation window ($\sim 50\ s$). This frame-wise averaging allows us to quantify the net propensity
of the bacterial population to move toward or away from the contact line at any given instant.

The results are summarized in Figure 6, where panels 6(a) and 6(c) display the temporal
evolution of the spatially averaged velocities $\bar{U}(t)$, and $\bar{V}(t)$, respectively. The corresponding
statistical distributions are shown in the form of box plots in Figures 6(b) and 6(d), enabling a
direct comparison between the live and dead bacterial suspensions. This combined analysis
provides insight into the magnitude of the cumulative transport behavior of the system, beyond the
instantaneous directional alignment explored earlier.

For $\bar{U}(t)$, the spatial (framewise) average of the ***x***-component of the velocity field, the live
bacterial samples exhibit a fluctuating profile centered near zero (see Figure 6 (a-b)), with a time-
averaged value of $\langle \bar{U}(t) \rangle = 0.007 \pm 0.868\ \mu\mathrm{m/s}$. Here the $\pm$ value denotes the temporal standard
deviation of the spatial mean velocity, $\sigma_{\bar{U}}$. This behaviour indicates the absence of a persistent
directional preference and frequent reversals in net flow. In contrast, the dead droplets display a
significantly higher and much sustained (noticeably less fluctuations) positive velocity, with a
mean of $0.729 \pm 0.311\ \mu\mathrm{m/s}$, reflecting a steady capillarity-driven flow toward the contact line.
This difference is statistically highly significant, as confirmed by a Welch's two-sample t-test ($t =$
$-8.34,\ p = 1.49 \times 10^{-13}$) and a Wilcoxon rank-sum test ($p = 1.45 \times 10^{-14}$), with a large
effect size (Cohen's d = $-1.29$).

For the spatially averaged $y$ velocity component, $\bar{V}(t)$, where negative values indicate
motion toward the contact line (as per our coordinate convention), a similar trend is observed,
though less pronounced (see Figure 6(c-d)). The dead bacterial droplets again exhibit a stronger
net flow toward the TCL, with a mean of $-0.692 \pm 0.338\ \mu m/s$, compared to $-0.521 \pm$
$0.834\ \mu m/s$ for the live case. Although the difference in means is smaller than for $\bar{U}(t)$, both the
t-test ($t = 2.04$, $p = 0.0431$) and Wilcoxon rank-sum test ($p = 0.00667$) indicate statistical
significance, with a modest effect size (Cohen's d = 0.31).

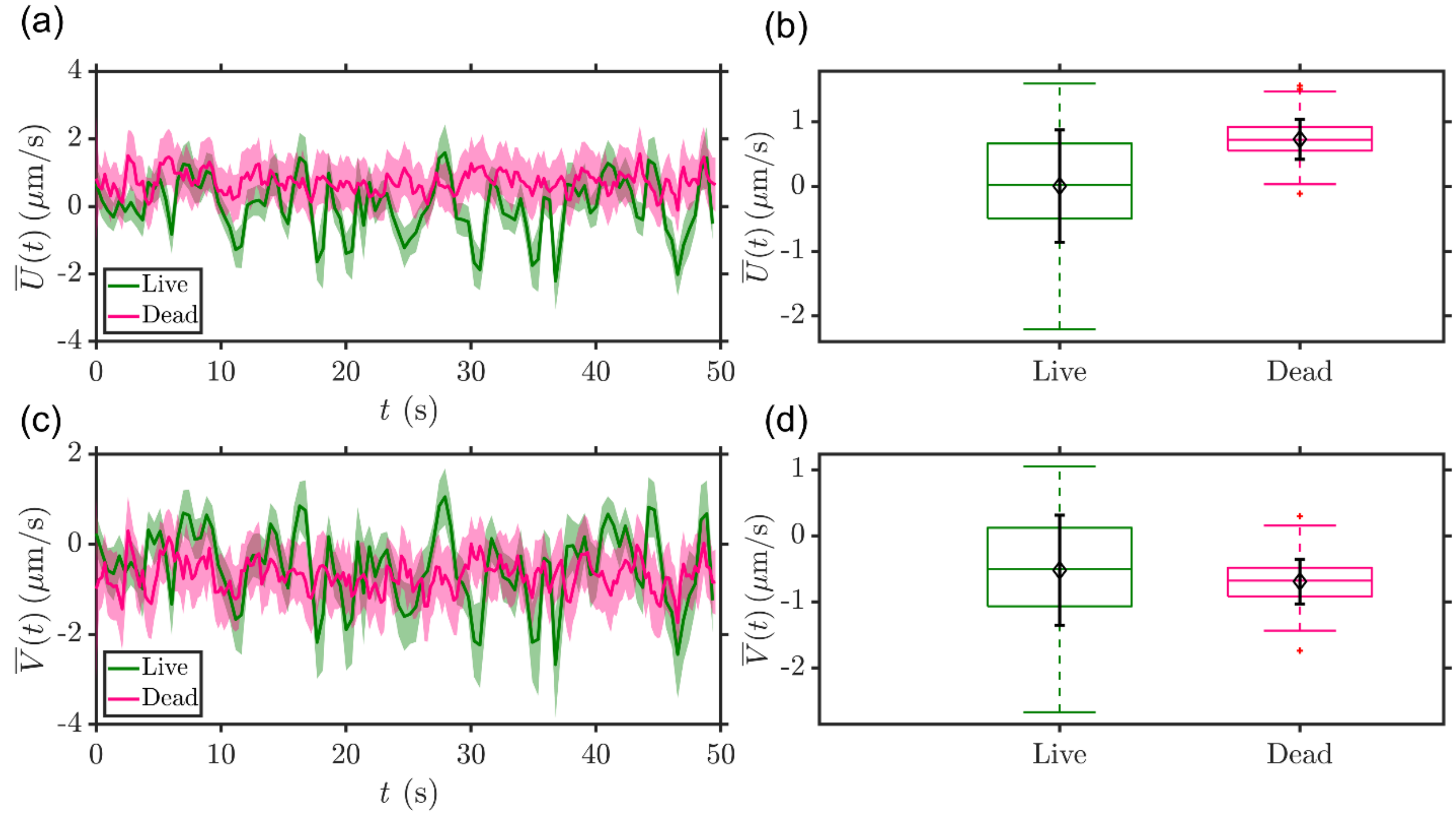

**Figure 6. Temporal statistics of spatially averaged velocity components in live and dead *E. coli*-laden droplets.** (a) Time series of the frame-wise spatial mean velocity, $\overline{U}(t)$, for live (green) and dead (pink) samples. The semi-transparent band around each curve shows the instantaneous spatial standard deviation, $\sigma_U(t)$, i.e., the intraframe (spatial) heterogeneity of the flow field. (b) Box plot summarizing the temporal distribution of $\overline{U}(t)$ over 50 $s$ observation window. The box spans the interquartile range, IQR ($25^{th}$ - $75^{th}$ percentiles), whiskers extend to the most extreme data points within 1.5 * IQR, and outliers are marked individually. The hollow black diamond denotes the time-average $\langle \overline{U}\ (t) \rangle$, and the vertical, solid black capped line at each group represents the temporal standard deviation, $\sigma_{\overline{U}}$ on each side (interframe variability of the spatial mean). (c) Analogous time series of the spatial mean $\overline{V}(t)$, with its intraframe spatial standard deviation $\sigma_V(t)$ shaded using semi-transparent band. Negative $\overline{V}$ values indicate net motion toward the contact line, as per our coordinate convention. (d) Boxplot summarizing the temporal distribution of $\overline{V}(t)$ over the 50 $s$ observation window with the same notational conventions as in (b): the box represents the IQR, whiskers extend to the most extreme data points within 1.5*IQR, the hollow diamond denotes $\langle \overline{V}\ (t) \rangle$, and the capped, black vertical line represents $\sigma_{\overline{V}}$ .

Beyond the trends in net transport, Figure 6 also provides insight into the nature of spatial
and temporal variability in the velocity field. In the time series plots (Figure 6 (a),(c)), the semi-
transparent shaded bands reflect the instantaneous spatial standard deviation ($\sigma_U(t)$, $\sigma_V(t)$),
capturing the degree of heterogeneity across the spatial domain at each time point. Notably, these
bands are consistently narrower for live samples than for dead ones, suggesting that live bacterial
flows, despite their temporal fluctuations, maintain tighter spatial coherence within each frame. In
contrast, the box plots (Figure 6 (b),(d)), which summarize the distribution of spatial mean
velocities over the 50 s window, show a higher temporal standard deviation for live samples, as
reflected by the taller capped $\sigma_{\bar{U}}, \sigma_{\bar{V}}$ bars around the time-averaged means ($\langle \bar{U}(t) \rangle$, $\langle \bar{V}(t) \rangle$). This
contrast implies that while the overall direction of motion in live samples fluctuates more
erratically over time, the velocity field at any given instant is spatially more ordered. In other
words, motile bacteria undergo frequent directional reversals, yet do so in a coordinated manner,
maintaining narrower local velocity distributions. This reduced intraframe variability suggests that
active bacteria reorganize collectively and synchronously, giving rise to coherent spatial coherence
despite dynamic temporal fluctuations.
Taken together, these results reinforce the notion that live bacterial populations resist the
imposed capillarity-driven flow through collective rheotactic behavior. While their net transport
toward the contact line is weaker and more variable than that of dead suspensions, they exhibit
reduced instantaneous spatial disorder, indicative of self-organization within the evolving velocity
field. We examine this spatiotemporal organization more quantitatively next by analyzing the
velocity-velocity correlation of the flow field.

**4.4.3. Spatial Velocity Correlations and Flow Coherence**

To evaluate the spatial coherence of the measured velocity field, we computed a two-point
Pearson correlation between the velocity-component time series measured at pairs of spatially
separated locations.This probes how temporally resolved velocity fluctuations at a given spatial
location are correlated with those at a neighboring location, separated by a known distance along
a given direction. Unlike instantaneous field-wise correlation schemes, this method captures the
temporal coherence of motion between adjacent spatial regions.
We use the generic notation $U_i(\mathbf{x}, t)$ to denote $x$ and $y$ components ($U$ and $V$, respectively)
of the velocity field, at a spatial location $\mathbf{x}$ and time $t$. $i \in \{x, y\}$. The pairwise correlation of the $i^{\text{th}}$ velocity component between locations $\mathbf{x}$ and $\mathbf{x} + s_j\hat{\boldsymbol{e}}_{\boldsymbol{j}}$, is computed using the Pearson
correlation[55] formulation as:

$$r_i\left(\mathbf{x}, \mathbf{x} + s_j\hat{\boldsymbol{e}}_{\boldsymbol{j}}\right) = \frac{\sum_{t=1}^{T}\left(U_i(\mathbf{x}, t) - \langle U_i(\mathbf{x})\rangle\right)\left(U_i\left(\mathbf{x} + s_j\hat{\boldsymbol{e}}_{\boldsymbol{j}}, t\right) - \langle U_i(\mathbf{x} + s_j\hat{\boldsymbol{e}}_{\boldsymbol{j}})\rangle\right)}{\sqrt{\sum_{t=1}^{T}\left(U_i(\mathbf{x}, t) - \langle U_i(\mathbf{x})\rangle\right)^2}\ \sqrt{\sum_{t=1}^{T}\left(U_i\left(\mathbf{x} + s_j\hat{\boldsymbol{e}}_{\boldsymbol{j}}, t\right) - \langle U_i(\mathbf{x} + s_j\hat{\boldsymbol{e}}_{\boldsymbol{j}})\rangle\right)^2}} \qquad \text{Eq. (2)}$$

Here, $\langle U_i(\mathbf{x})\rangle$ represents the temporal mean of the $i^{\text{th}}$ velocity component at the location $\mathbf{x}$,
computed over the entire time window $T$. The separation distance $s_j$ is taken along the unit vector
direction $\hat{\boldsymbol{e}}_{\boldsymbol{j}} \in \{\hat{\boldsymbol{e}}_{\boldsymbol{x}}, \hat{\boldsymbol{e}}_{\boldsymbol{y}}\}$, where $j \in \{x, y\}$.
The mean spatial correlation, $C_{U_i}^{temporal}$ is then obtained as a function of $s_j$ by averaging the
pairwise coefficients over all valid spatial pairs at separation $s_j$ as

$$C_{U_i}^{temporal}\left(s_j\right) = \frac{1}{N_s}\sum_{\mathbf{x}} r_i(\mathbf{x}, \mathbf{x} + s_j\hat{\boldsymbol{e}}_{\boldsymbol{j}})$$

where $N_s$ is the number of valid spatial pairs with separation $s_j$ along direction $\hat{\boldsymbol{e}}_{\boldsymbol{j}}$.
As no prominent directional preference/tendency was observed thus far in the analyzed flow field,
the correlation is independently evaluated for both the $U$ and $V$ velocity components along both $x$
and $y$ directions. The resulting correlation profiles are presented in Figure 7: panel (a) shows
$C_U^{temporal}(s_x)$, (b) shows $C_U^{temporal}(s_y)$, (c) shows $C_V^{temporal}(s_x)$ and (d) shows $C_V^{temporal}(s_y)$.
Each panel represents the spatially averaged correlation coefficient as a function of separation
distance for the respective velocity component-direction pairs, computed independently for four
experimental conditions, namely live samples at concentrations $\approx 10^7$ CFU/ml and $\approx 10^6$ CFU/
ml, and dead samples at concentrations $\approx 10^7$ CFU/ml and $\approx 10^6$ CFU/ml, allowing direct
comparison of spatial correlation behavior across motility and concentration.
As seen in Figure 7, live bacterial samples consistently exhibit stronger correlation
magnitudes that decay more gradually with distance. For example, live samples with estimated
concentration $\approx 10^7$ CFU/ml show peak correlations in the range $0.955 - 0.961$ and do not decay
below $1/e$ of their peak value within the measurement window ($\sim$200 μm), for any of the four
velocity component ($U$ or $V$)- correlation direction ($x$ or $y$) pairs, indicating strong long-range coherence in the flow field (see Figure 7 (a-d)). Similar long-range coherence is observed in the
live samples with concentration $\approx 10^6$ CFU/ml, for both the velocity components when the
correlation is computed in the $x$ direction. Even for correlation along $y$-direction, although overall
peak correlations are slightly reduced $(0.912 - 0.921)$, the correlation length (i.e., the separation
length over which the correlation decays below $1/_e$ times the peak correlation) extends up to
$\sim 150 - 160$ μm in the $y$-direction, again highlighting coordinated motion.

In contrast, dead bacterial samples exhibit lower peak correlations $(0.844 - 0.888)$ and
much shorter correlation lengths ($\sim 40$ μm across all four directions and velocity component pairs),
reflecting a rapid loss of coherence. This is consistent with passive, capillarity-driven transport in
the absence of active, coordinated motility.

A log-linear fit-based decay analysis further supports these trends. The initial decay slopes
are substantially flatter in live cases $(-0.002$ to $-0.008)$ compared to dead cases $(-0.015$ to $-$
$0.031)$, quantitatively confirming the presence of long-range order in motile suspensions vs short-
range, incoherent motion in passive, non-motile controls.

Furthermore, a clear concentration dependence is observed in the live samples. The lower
concentration case $\approx 10^6$ CFU/ml shows reduced peak correlation and shortened correlation
length compared to the $\approx 10^7$ CFU/ml case, highlighting the role of bacterial concentration in
enhancing collective motion. Dead samples, by contrast, show much less pronounced sensitivity
to concentration, consistent with their passive motion by evaporation-driven capillary flow.

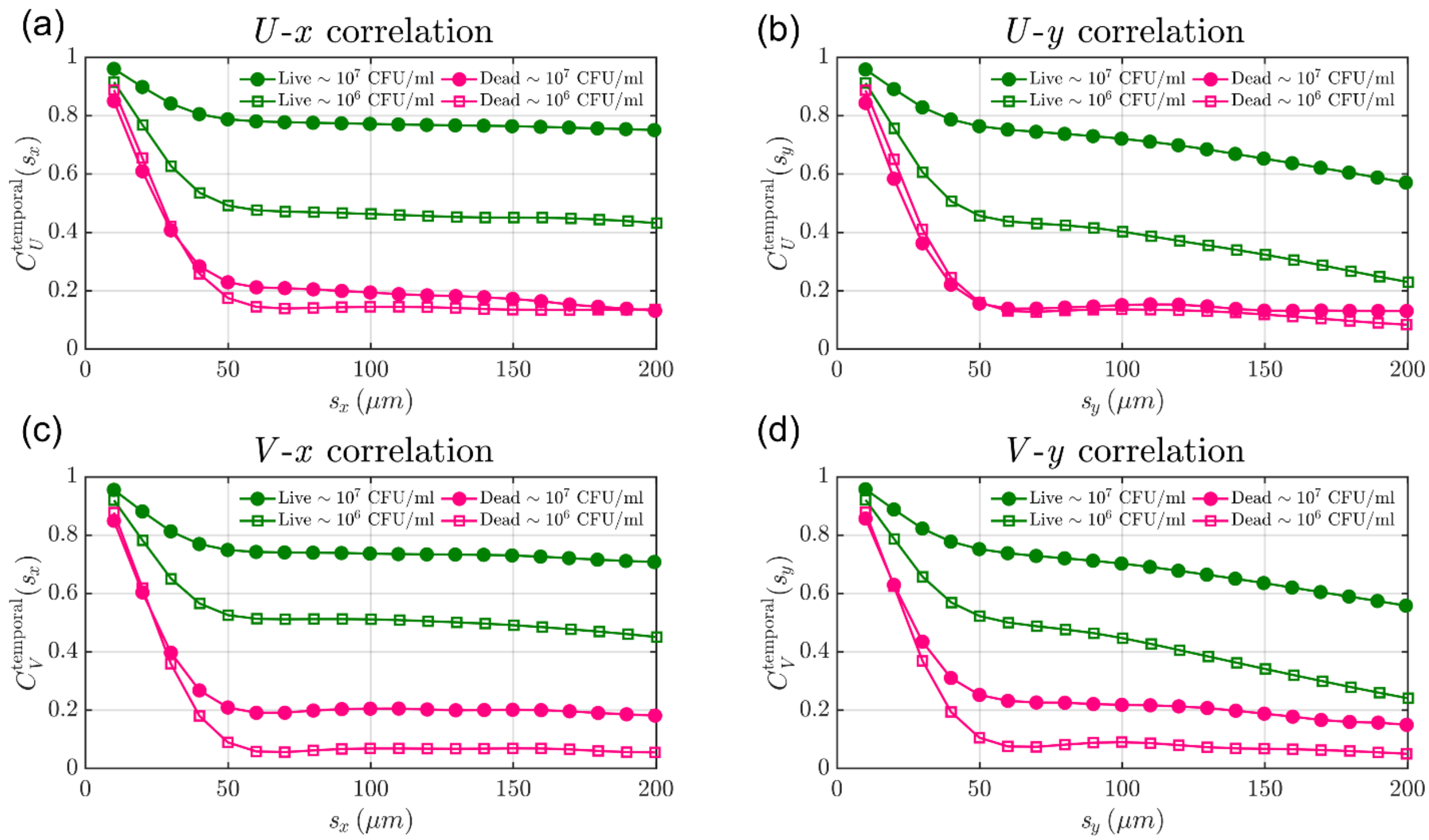

**Figure 7: Spatial velocity-velocity correlation profiles computed using temporally resolved**
**two-point Pearson correlation analysis.** Each panel presents the correlation coefficient
$C_{U_i}^{temporal}(s_j)$ as a function of spatial separation $s_j$, where $U_i \in \{U, V\}$ denotes the velocity
component and $j \in \{x, y\}$ indicates the direction of spatial separation. Specifically, (a) shows
correlations of the $U$-component along the $x$-direction, (b) $U$ along $y$, (c) $V$ along $x$, and (d) $V$
along $y$. Live bacterial samples consistently exhibit higher peak correlations and slower decay,
indicating long-range coherence in the flow field, whereas dead samples display rapid
decorrelation over short distances. Color and marker schemes denote bacterial motility (live vs.
dead) and concentration levels as indicated in the legend.

Note that the bacterial concentrations used in the present study are lower than those
typically associated with dense active-suspension collective behavior[56]. Therefore, the enhanced
spatial correlations observed for live *E. coli* suspensions should not be treated as synonymous with
classical bulk collective behavior reported in dense bacterial suspensions. Rather, the higher peak
correlation, slower correlation decay, and longer correlation length observed here indicate stronger
spatial coherence in the measured velocity field of live suspensions compared with dead
suspensions. This suggests motility-mediated active coordination within the droplet, without implying the fully developed collective state reported for dense bacterial suspensions, such as
chaotic turbulent-like flow behavior, large-scale vortices, or persistent swirling patterns.

As discussed earlier, live bacteria oppose capillary flow and, at higher concentrations,
exhibit collective motion. Here, we isolate and discuss the physical aspects underlying each
phenomenon. First, we focus on the question: *why do live bacteria oppose capillary flow*?
Additionally, *why is the velocity of the live bacterial flow field near the contact line lower*
*compared to that of dead bacteria*? To put this into perspective, the resistance of bacterial motion
to the prevailing capillary flow implies that live bacteria tend to move upstream. However, the
dominance of capillary motion ultimately drags or attempts to drag the bacteria toward the contact
line. This competition between capillary flow and the upstream motion of live bacteria results in a
lower velocity for the live bacterial flow field compared to that of dead bacteria (see Figures 4 (a-
c)). This leads to a critical question: *why do live bacteria exhibit upstream motion*?

In this context, we should note here that motile bacteria tend to exhibit upstream motion at
the liquid-solid interface due to positive rheotaxis[37–39]. Rheotaxis is a survival mechanism that
allows bacteria to move towards nutrient-rich zones or away from harmful conditions in a dynamic,
flowing environment. In rheotaxis, motile bacteria like *E. coli* reorient themselves against a
unidirectional flow. This behavior arises from the interplay between velocity gradients and the
helical shape of the flagella, which together generate a torque that alters the swimming direction
of the bacteria. The chiral shape of the flagella ensures the presence of a lift component acting on
the moving flagella, while the bacterial body, being non-chiral, does not experience any lift force.
Instead, a drag force acts on the bacterial body, anchoring the overall motion. The combination of
drag and lift forces generates a torque that opposes the flow direction.

This work does not aim to separately investigate the phenomena of rheotaxis or collective
migration but instead uses these phenomena as frameworks to explain the anomalous macroscopic
surface interaction behavior demonstrated by bacteria-laden droplets. Furthermore, it is important
to distinguish between bacterial adhesion/attachment to surfaces at the individual scale and the
force of adhesion of a bacteria-laden droplet. A bacteria-laden droplet may exhibit low adhesion
force with a substrate at an early timescale, but this does not imply that bacteria themselves are
non-adhesive to the surface. The adhesion force of a bacteria-laden droplet is a macroscopic
quantity that involves contributions from pinning effects and the interfacial energy of the droplet,
whereas the adhesion of individual bacteria to surfaces is a mechanobiological phenomenon where bacteria form appendages to remain adherent to surfaces and form biofilms over longer timescales as a survival mechanism. These are two vastly different phenomena occurring at entirely different timescales. The reduced adhesion of bacteria-laden droplets at the early stage does not defy the conventional wisdom that bacteria prefer to remain adherent to surfaces and form biofilms. Capillary flow induces recirculation of bacterial cells, which tends to carry bacteria away from the surface. Bacteria's attempt to resist capillary flow via positive rheotaxis results in two outcomes at different timescales-short-term reduction of the contact line pinning force, detectable via early-stage adhesion measurement, and long-term reduction of overall recirculation, which facilitates the formation of appendages and enhances bacterial adhesion, ultimately leading to biofilm formation[3].

### 4.5. Sliding Behavior Explains Near Contact Line Dynamics of Bacterial Droplets:

It is evident from Figure 2 that the wetting signature of *E. coli*-laden droplets on superhydrophobic surfaces is insufficient to capture the true nature of surface interactions. Optical goniometry measurements on these surfaces fail to distinguish between live and dead bacterial suspensions and do not reveal consistent trends with varying concentration. As discussed earlier, such inconclusive wetting behavior is expected on superhydrophobic substrates, primarily due to challenges in accurately resolving the baseline of the droplet and the associated systematic uncertainties[53] caused by light scattering and diffraction near the triple contact line. In contrast, cantilever-based adhesion measurements[47–51,57] performed on the same superhydrophobic surface demonstrate a clear and consistent difference between live and dead droplets and reveal systematic trends with concentration.

In order to pinpoint whether this difference arises from the intrinsic nature of the bacterial load or is simply a manifestation of the wettability of the test substrate, a logical follow-up is to probe both the dynamic wetting behavior and the adhesion signature on a moderate-wettability surface. It could be anticipated that if there is a true physical difference in the nature of surface interaction that is not substrate-dependent, it should be reflected in contact angle measurements on moderate-wettability surfaces, which are less susceptible to systematic optical errors. However, direct adhesion quantification using cantilever probes on moderate-wettability surfaces is experimentally challenging, as the probe droplet often fails to detach from the substrate. To circumvent this, we devised an alternative sliding experiment on moderate wettability polymethyl methacrylate
(PMMA) (static water contact angle ~70°)[50] to probe the dynamic contact line behavior of live and
dead bacterial droplets induced by controlled surface tilt.

In this protocol, a bacterial droplet is gently deposited on the PMMA surface mounted on the tilting
stage of a contact angle goniometer, which is then tilted slowly at a controlled rate of ~10°/min.
The evolution of droplet shape is recorded using a CMOS camera integrated with the goniometer.
This setup enables the simultaneous measurement of the evolution of the contact angles of the
leading edge (right edge, CA(r)) and the trailing edge (left edge, CA(l)), as well as the projected
contact diameter. By continuously tracking the contact angles throughout the tilting process, we
capture the dynamic variation of the contact line under a gradually applied external force (gravity).
Additionally, by identifying the point at which the droplet first initiates sliding, the corresponding
leading and trailing edge contact angles at that moment provide the advancing contact angle (ACA)
and receding contact angle (RCA), respectively.

Dynamic contact angle measurements revealed a counterintuitive trend: while the ACA remains
comparable between live ($86.4 \pm 0.5°$) and dead droplets ($89.5 \pm 3.9°$), the RCA is significantly
lower for live droplets ($32 \pm 0.6°$) than for dead ones ($49.3 \pm 3.7°$), resulting in a higher contact
angle hysteresis (CAH) in live samples. At first glance, this would suggest that dead droplets
should have higher contact line mobility and should slide more easily, as a lower RCA is often
associated with greater pinning and energy dissipation. However, the comparative sliding
dynamics between live versus dead droplets, visualized in Supporting Video S5, revealed that live
droplets initiate sliding motion at noticeably lower tilt angles and exhibit more frequent depinning
events, suggesting enhanced contact line mobility.

Figure 8 provides the motion analysis of the edges of the droplets in greater detail (also see
Supporting Video S5). A few distinct trends differentiate the sliding behavior of live (Figure 8a)
and dead *E. coli*-laden droplets (Figure 8b). Live bacterial droplets display frequent stick-slip
behavior of the contact line. At a tilt angle as low as ~5.3°, the leading edge of the droplet depins
and moves forward, followed by several subsequent depinning events, as indicated by the step-
like increments in the projected base diameter shown in Figure 8a. The trailing edge remains
pinned until ~36° tilt angle, beyond which both edges become mobile and the entire droplet starts to slide down. The onset of sliding is marked by a blue circle in CA(l) vs tilt angle plot in Figure 8a, where the contact angle at the just mobile trailing edge exhibits a near-constant contact angle. Beyond this point, a sharp elongation of the footprint of the droplet can be observed, with the contact diameter increasing from approximately 8.1 mm to 10 mm. This can be attributed to the combined effects of the relatively lower surface tension of the live bacterial droplet and the increased inertia at higher tilt angles. In general, the live bacterial droplets exhibit a higher propensity for elongation in footprint, with the diameter increasing from ~6 mm to ~10 mm over the course of the sliding experiment.

In contrast, for dead bacterial droplets (Figure 8b), both the leading and trailing edges remain pinned over a noticeably wider range of tilt angles. The first depinning event of the leading edge (right edge) occurs only at around ~22.4°, while the depinning of both edges/onset of sliding occurs at ~44° tilt. Furthermore, the contact diameter grows more gradually, from ~5.95 mm to ~7.1 mm, with significantly fewer depinning events observed. While the elongation of contact diameter for live droplets could possibly be attributed to their lower surface tension, it is important to emphasize that the frequent stick-slip behavior, characterized by depinning of the contact line, clearly points to enhanced contact line mobility in live bacterial suspensions.

Despite having a lower RCA, live droplets demonstrate significantly greater contact line mobility and slide more readily than dead droplets. This counterintuitive behavior suggests that macroscopic goniometric indicators such as RCA and CAH alone are insufficient to describe the dynamic wetting/surface interaction behavior of biologically active droplets. The enhanced mobility of live droplets likely arises from internal activity, such as bacterial motility, intrinsic tendency to oppose capillary flow, upstream migration behavior, and inter-bacterial interactions, all of which facilitate weakening of pinning sites and promote coordinated depinning events.

At this juncture, a few fundamental questions naturally arise: *Can the observed differences in sliding behavior between live and dead bacterial suspensions be attributed simply to their difference in surface tension? Further, can we mimic similar behavior by replacing live bacteria with inert, similarly sized microparticles at comparable concentrations while maintaining a similar surface tension to that of the live bacterial suspension? In such a case, can we still expect* *enhanced contact line mobility and stick-slip behavior?* To address these questions, we designed two comparative proxy systems and studied their sliding behavior.

In the first system, we focused on isolating the effect of surface tension. We compared the sliding behavior of DI water droplets (surface tension $\approx$ 72 mN/m) with that of a binary mixture of ethylene glycol (EG) and DI water, where the mole fraction of EG, $x_{EG} = 0.1$, resulting in a surface tension of ≈62.29 mN/m. The sliding dynamics for this system are presented in Figure S6(a-b) of the supporting information, with comparative sliding videos provided in Supporting Video S6.

In the second system, we aimed to mimic the particle effect. We doped the aforementioned two solutions (DI water and EG-water mixture) with 1 μm-sized polystyrene microparticles, maintaining a doping concentration of $\approx 10^7$ particles/mL to match both the size and concentration of live bacteria in the test suspensions. Further details regarding the particles used and the preparation protocol are provided in the Methods section. This setup allowed us to study the influence of inert microparticles on the sliding behavior and to isolate the role of active bacterial motility. Figure S6(c-d) of the supporting information captures the comparative sliding dynamics of this system.

As shown in Figure S6 and Supporting Videos S6 and S7, neither the low surface tension binary mixture alone nor the corresponding microparticle-laden suspensions could replicate the enhanced contact line mobility and dynamic stick-slip behavior observed for live *E. coli*-laden droplets (Figure 8a and Supporting Video S5). These results strongly suggest that the distinctive sliding behavior of live bacterial droplets cannot be attributed solely to reduced surface tension or the presence of similarly sized microparticles; rather, it points towards an intrinsic contribution from active bacterial motility and collective behavior.

Together, these findings emphasize that active internal bacterial dynamics, rather than macroscopic wetting signatures obtained from goniometry, govern the near-contact line behavior and the resulting surface interaction behavior of live bacterial droplets. It leads to fundamentally different surface interactions compared to inert (dead) systems and has important implications for how bacterial suspensions interact with coatings, medical devices, or host surfaces in real-world scenarios.

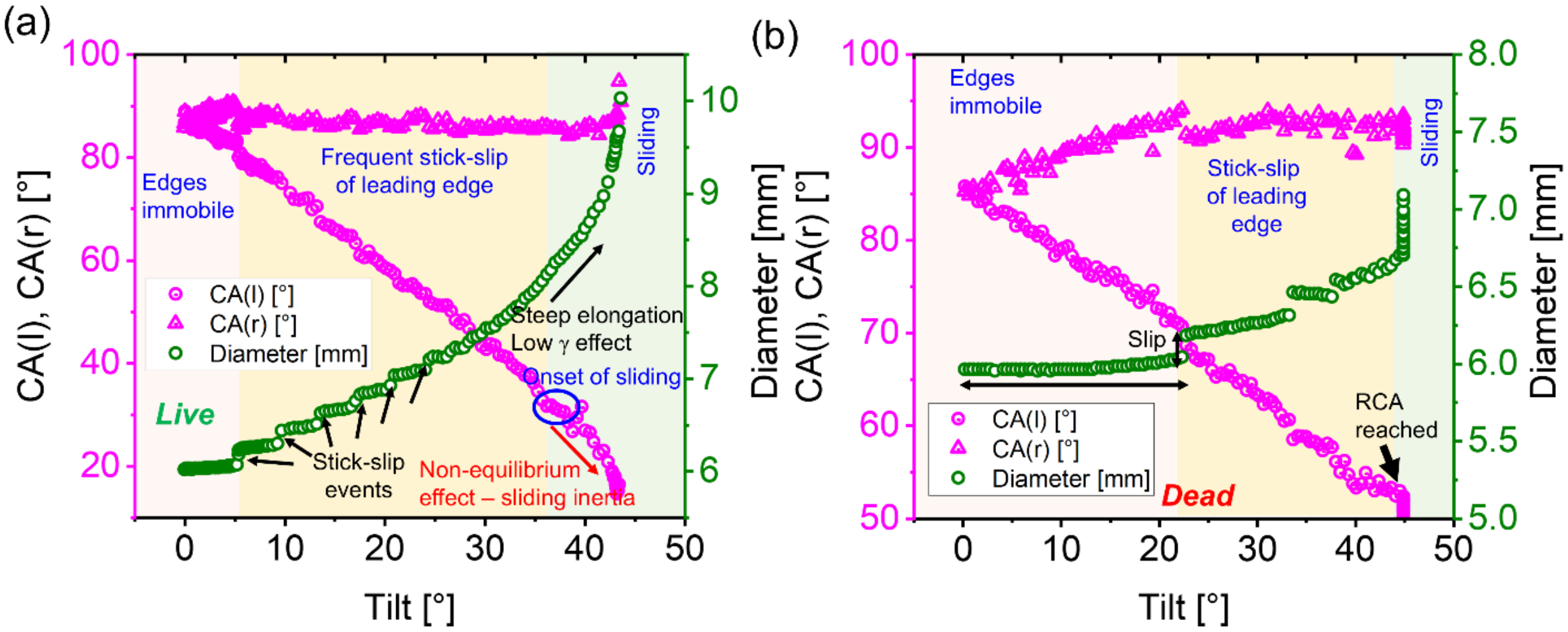

**Figure 8: Sliding behavior of live and dead *E. coli*-laden droplets on a moderate wettability PMMA surface.** (a-b) Tilt-induced sliding behavior captured through the evolution of the contact angle of the leading edge (CA(r)) and the trailing edge contact angle (CA(l)), and the projected (contact) diameter as a function of tilt angle for (a) live and (b) dead bacteria laden droplets.

### 4.6. Study Scope and Limitations

This study focuses on the early-stage ($\sim O(100\ s)$) interfacial interaction of *E. coli*-laden droplets with solid surfaces, specifically quantifying the surface adhesion (i.e., the force required to detach the entire droplet from the test surface), rather than investigating bacterial attachment at the single-cell level. We do not examine reversible/irreversible attachment, receptor-specific binding, or long-term biofilm development. Instead, our objective is to understand how flagellar motility modulates droplet-scale adhesion, which has practical implications for the removal of contaminated droplets from surfaces.

While extracellular polymeric substance (EPS) secretion can begin within minutes[58–60] of bacterial contact with a surface and is known to enhance adhesion by forming a viscoelastic matrix, our measurements occur on timescales that are still below the earliest reported onset of EPS activity. For example, EPS-associated gene expression has been observed as early as $\approx 5$ min[59] in *Pseudomonas aeruginosa*, whereas all our wetting, adhesion force measurement, or $\mu-$PIV experiments are completed within $\approx 100\ s$ of droplet-surface contact. At these early timescales, whether secreted matrix components exert a measurable influence on contact line behaviour cannot be determined with certainty. The possibility of EPS production was not accounted for in our analysis. Nonetheless, since EPS is generally associated with increased adhesion, its presence would be expected to oppose rather than amplify the contact line mobility observed in live droplets. Live droplets in our experiments consistently show reduced adhesion compared to dead counterparts: a trend that becomes more pronounced at higher bacterial concentrations. Thus, even if EPS secretion contributes to the interfacial behavior within our measurement window, logically, it would further reinforce our conclusion that active motility plays the dominant role in facilitating contact line mobility during the early stages of droplet-surface interaction.

We also note that percentage viability quantification (i.e., determining the fraction of live versus dead cells) in the live bacterial suspensions was not performed in this study. While qualitative assessments using motility agar, fluorescence microscopy, and TEM imaging confirmed the presence of motile, flagellated cells, future work could include systematic viability staining and cell counting to refine the quantitative interpretation of motility-driven effects. Further, 70% IPA treatment used for preparing dead bacterial samples may introduce lysis products into the suspension. While centrifugation at 10,000 rpm followed by removal of the supernatant, as employed during sample preparation, likely eliminates smaller cellular residues, no dedicated steps were taken to remove larger lysis byproducts or membrane fragments. Dynamic light scattering (DLS) indicated the presence of potential agglomerates in the dead samples, which could have some influence on the interfacial behavior. Future work could incorporate additional purification protocols or explore alternative inactivation methods, such as heat killing, to better isolate the effects of bacterial viability. Future studies involving cell-free supernatant controls, independent surface-charge measurements, and quantitative comparison of live vs. dead cell-surface morphology could provide additional biological insight by decoupling biochemical and cell-surface contributions from motility-mediated effects.

Finally, this study is restricted to a single motile species, *Escherichia coli* K-12, which exhibits flagella-driven run-and-tumble dynamics. The findings may not be generalizable to non-motile strains or bacteria employing alternative modes of motility (e.g., pili-mediated twitching or gliding). All adhesion experiments were performed on a single type of superhydrophobic surface;

variations in surface chemistry or topography were not explored and remain beyond the present scope.

## 5. CONCLUSION

This study investigated how bacterial motility modulates the wetting and interfacial behavior of sessile, *E. coli*-laden droplets, focusing on early-stage surface adhesion. For wetting characterization, we used contact angle goniometry and a cantilever-deflection-based adhesion measurement system. Unfortunately, the dynamic wetting characterization on superhydrophobic surfaces remained inconclusive, as no definitive or statistically significant trends were observed either between live and dead *E. coli*-laden droplets or in their concentration dependence. However, switching to a more direct surface interaction quantification method, the cantilever deflection method, enabled us to identify several interesting trends. Despite their lower surface tension, the adhesion force of live *E. coli*-laden droplets (concentration $\approx 10^7$ CFU/ml) was significantly lower (~9 μN) compared to ~39 μN for the corresponding dead droplet case. Moreover, the adhesion force for live bacterial droplets decreased with increasing bacterial concentration, while for the dead bacteria, adhesion force underwent slight increase with concentration, as expected.

To probe the physical mechanisms behind this anomalous adhesion signature, we visualized the internal flow fields within sessile droplets using $\mu-$PIV and particle tracking. These measurements revealed that the internal motion of live bacteria differs fundamentally from that of dead bacteria. While dead bacteria are passively advected by evaporation-driven capillary flow toward the contact line, live bacteria actively resist this transport. This resistance manifests as lower net velocities toward the contact line, greater directional variability, and broader angular distributions of motion. Importantly, individual pathlines of live bacteria also showed frequent directional reversals and crossovers, indicating rheotactic behavior and inter-bacterial interactions.

Statistical quantification of flow directionality confirmed these differences: dead droplets exhibited high and stable directional alignment with capillary flow, while live droplets showed large temporal fluctuations, highlighting a dynamic competition between motility and imposed capillary flow. Yet, despite this temporal disorder, motile bacteria demonstrated remarkable spatial coherence. Correlation analyses of the velocity field revealed that live bacterial suspensions possess long-range spatial correlations, which decay gradually with distance, in contrast to the short-range, incoherent motion of dead bacteria. This spatial organization, noticeable even at moderate concentrations ($\approx 10^6$ CFU/ml), albeit to a lesser extent, suggests the presence of
collective behavior in motile suspensions. The emergence of long-range coherence and reduced
intraframe spatial variability in live samples points toward active coordination, potentially
mediated by rheotaxis or quorum-sensing mechanisms. We hypothesize that these coordinated
dynamics, coupled with reduced flow toward the contact line, disrupt the evaporation-driven
capillary transport typically responsible for reinforcing pinning. As a result, live *E. coli*-laden
droplets exhibit weaker contact line pinning tendency and lower net adhesion forces.

In addition to internal flow reorganization, sliding experiments on moderate-wettability
surfaces provided further evidence of motility-induced alteration of contact line dynamics. Despite
exhibiting a lower receding contact angle and higher hysteresis, typically associated with stronger
pinning, live bacterial droplets initiated sliding motion at lower tilt angles and displayed more
frequent depinning events than dead droplets. This counterintuitive behavior underscores the
insufficiency of macroscopic contact angle metrics in describing biologically active droplet
dynamics. The enhanced contact line mobility and stick-slip behavior observed in live droplets
could not be replicated by surface tension-matched or microparticle-doped proxy systems,
affirming the intrinsic role of motility. These results highlight how active bacterial motion, beyond
simply resisting internal flow alignment, can locally weaken pinning force and facilitate contact
line depinning, thereby reducing net adhesion and altering sliding dynamics.

Much of the existing literature in this regard has examined bacterial adhesion in the context
of long-term cell or biofilm attachment, governed by DLVO-type interactions[8,9], surface
chemistry[13,14], roughness[16], and/or EPS secretion[20–22]. In contrast, our study focuses on droplet-
scale adhesion, i.e., the force required to detach a bacteria-laden droplet from a surface, rather than
cell attachment itself. While previous works have explored evaporation-driven deposition and
contact line dynamics in bacterial droplets[23–25], the direct role of motility in modulating early-stage
adhesion has remained unexplored[26]. Here, we show that bacterial motility can contribute to the
reorganization of internal flow fields and weaken contact-line pinning, providing a mechanistic
interpretation of early-time-scale (~$O$ (100s)) droplet-surface interaction that differs from classical
passive wetting behavior.

In summary, our findings offer new insights into how microscale motility of bacterial
populations can reorganize internal flow fields collectively and reshape macroscopic interfacial
behavior, with implications for early-stage adhesion, droplet roll-off, contamination dynamics, and the design of self-cleaning or anti-biofouling surfaces. Future work could further decouple these multiscale interactions by tuning environmental stimuli or exploring different bacterial strains to better understand, control, and exploit motility-induced wetting transitions.

## 6. ACKNOWLEDGEMENT

S. K. Mitra acknowledges the funding from the Natural Sciences and Engineering Research Council (NSERC), Canada, in the form of the NSERC Alliance Grant Mission (ALLRP 570425-2021). S. Misra thankfully acknowledges funding support provided to him by Anusandhan National Research Foundation (ANRF), India in the form of Ramanujan Fellowship (RJF/2025/000356). The authors thank Mishi Groh (Department of Biology, University of Waterloo) for assistance with TEM imaging of *E.coli* and gratefully acknowledge Prof. Juewen Liu (Department of Chemistry, University of Waterloo) for providing access to the dynamic light scattering (DLS) facility. The authors also acknowledge the use of OpenAI's ChatGPT platform for proofreading and minor paraphrasing parts of the manuscript.

1277
1278
1279
1280
1281

# Supporting Information

**Motile *Escherichia coli*-laden Droplets Exhibit Reduced Adhesion and Anomalous Wetting Behavior**

Sirshendu Misra[±,a,b], Sudip Shyam[±,a], Priyam Chakraborty[a,c], Sushanta K. Mitra[a],*

[±]: *Authors contributed equally*

*: Corresponding author. Email: skmitra@uwaterloo.ca

[a] Micro & Nano-Scale Transport Laboratory, Waterloo Institute for Nanotechnology, Department of Mechanical and Mechatronics Engineering, University of Waterloo, 200 University Ave W, Waterloo, Ontario N2L 3G1, Canada

[b] Current Affiliation: Department of Mechanical Engineering, Indian Institute of Science, Bengaluru, 560012, India

[c] Current Affiliation: Department of Aerospace Engineering, Indian Institute of Technology Kharagpur, 721302, India

## 1. Supporting Videos

**Supporting Video S1:** Concentration dependence of surface adhesion of live bacteria-laden droplets

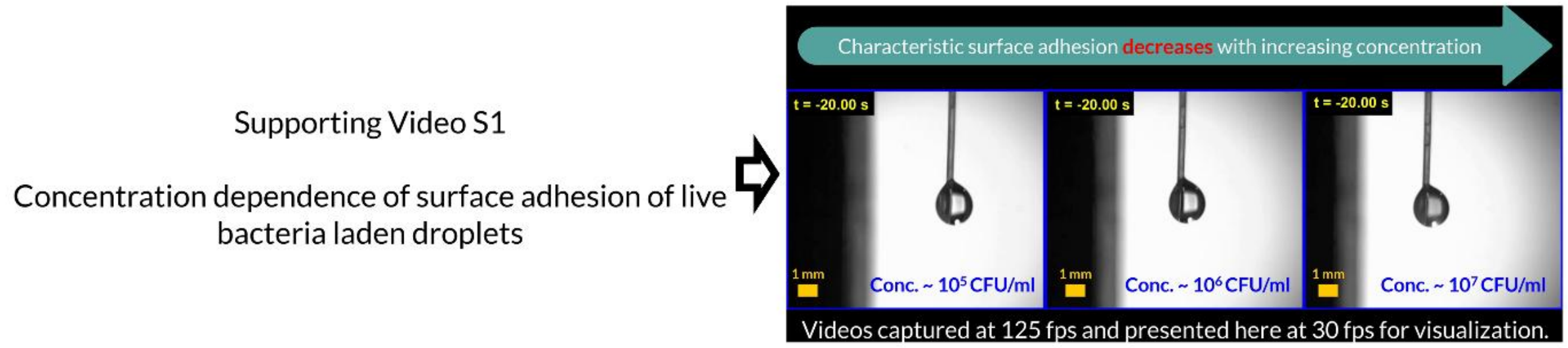

**Supporting Video S2:** Concentration dependence of surface adhesion of dead bacteria laden droplets

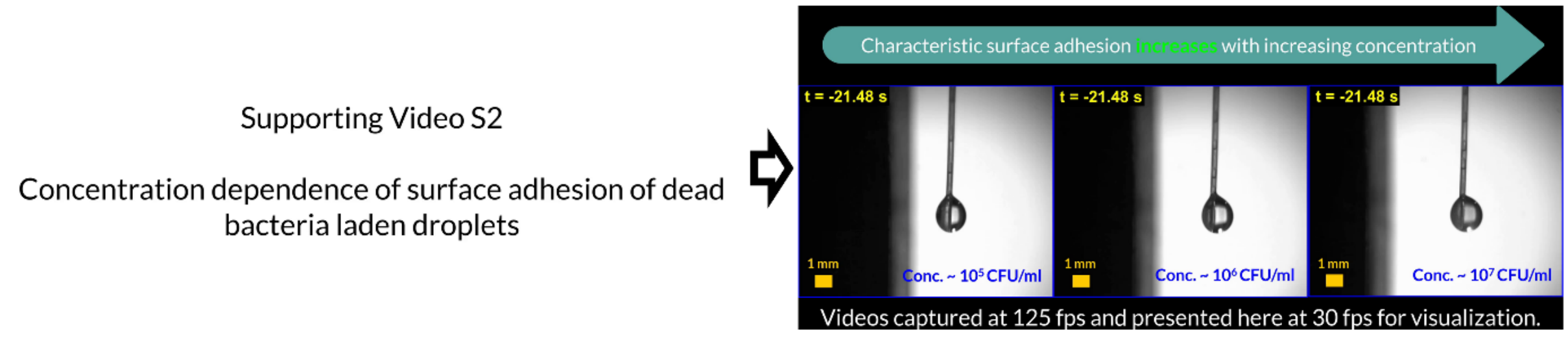

**Supporting Video S3:** Evolution of the velocity field of the live bacterial tracers within the droplet

**Left panel** illustrates the temporal evolution of the resultant velocity field with velocity vectors overlaid on a surface plot showing the velocity magnitude.
**Right panel** presents the local direction of $u$-velocity at the PIV grid points: red indicates positive $u$-velocity (flow towards the contact line), and blue indicates negative $u$-velocity (flow away from the contact line). Black regions mark the area outside the droplet, effectively denoting the contact line.

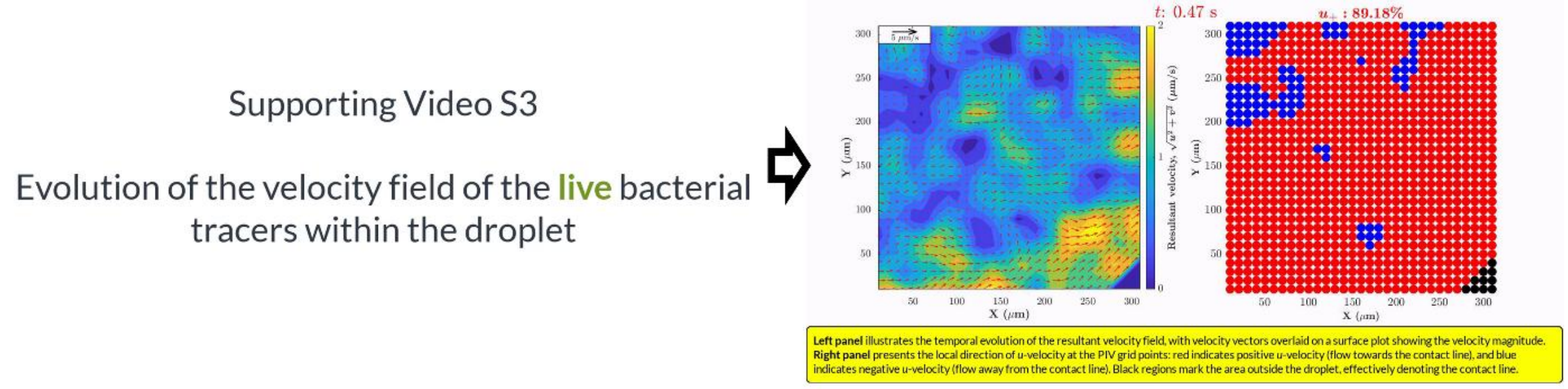

**Supporting Video S4:** Evolution of the velocity field of the dead bacterial tracers within the droplet

**Left panel** illustrates the temporal evolution of the resultant velocity field with velocity vectors overlaid on a surface plot showing the velocity magnitude.
**Right panel** presents the local direction of $u$-velocity at the PIV grid points: red indicates positive $u$-velocity (flow towards the contact line), and blue indicates negative $u$-velocity (flow away from the contact line). Black regions mark the area outside the droplet, effectively denoting the contact line.

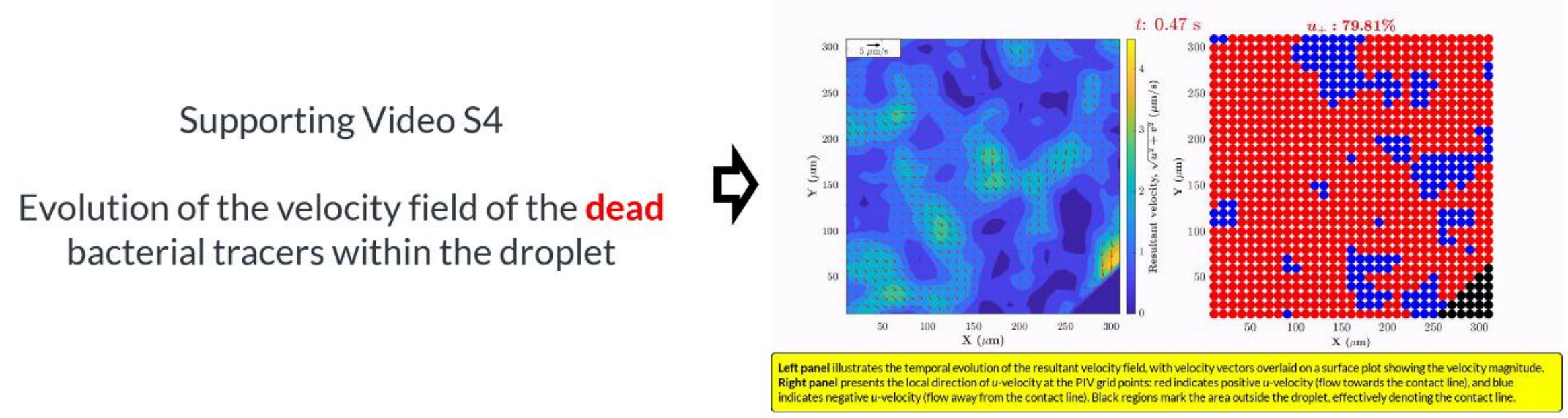

**Supporting Video S5:** Effect of bacterial motility on droplet sliding behavior: Comparative sliding dynamics of live vs dead bacteria-laden droplets

$CA_l$, $CA_r$ represent the contact angles at the trailing and leading edges, while $d_{base}$ denotes the projected contact diameter of the droplet. $\theta_{tilt}$ is the tilt angle with the horizontal plane.

Supporting Video S5:

Effect of bacterial motility on droplet sliding behavior:

Comparative sliding dynamics of live vs dead bacteria-laden droplets

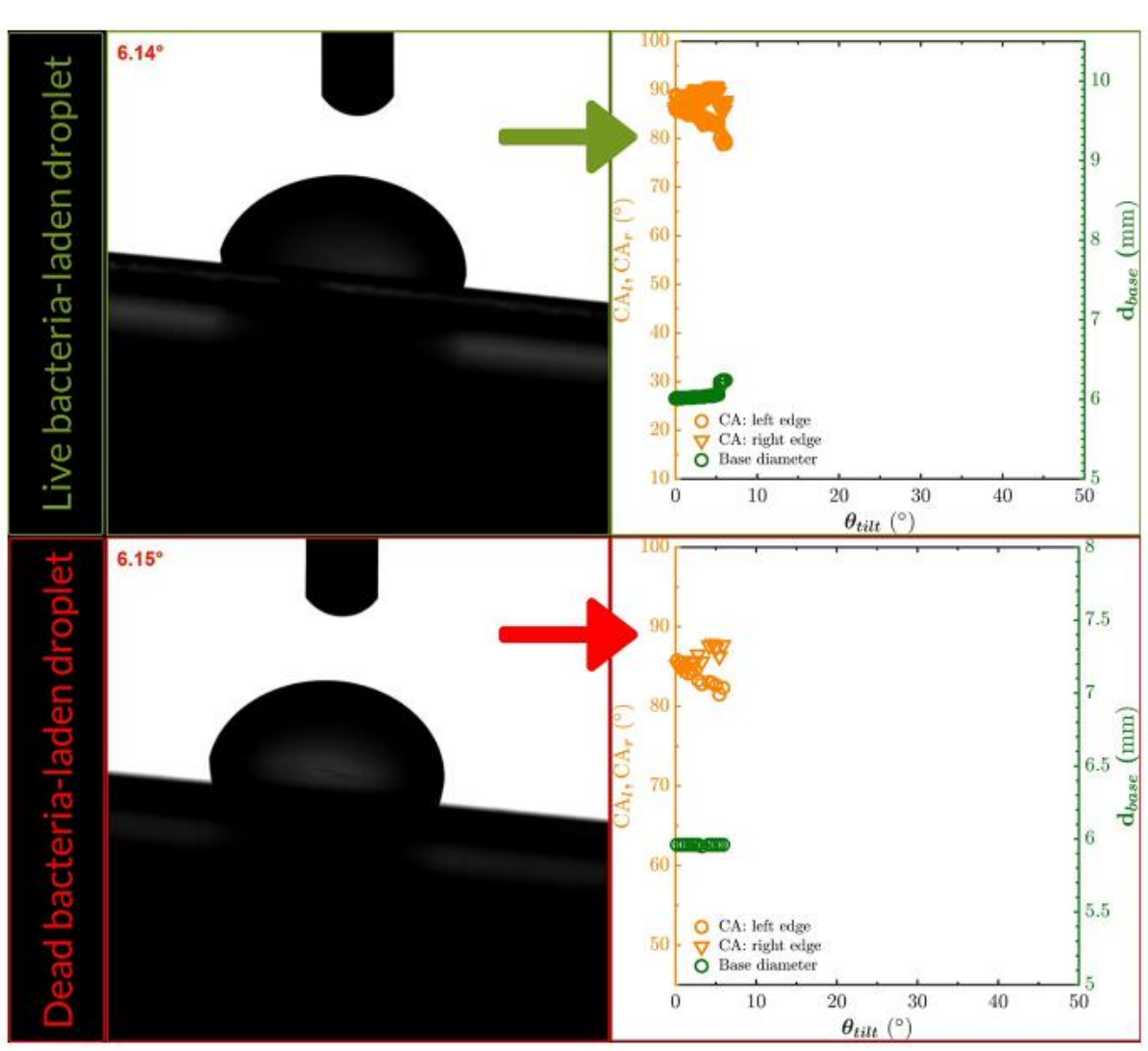

**Supporting Video S6:** Comparative effect of surface tension on the sliding dynamics of droplets $CA_l$, $CA_r$ represent the contact angles at the trailing and leading edges, while $d_{base}$ denotes the projected contact diameter of the droplet. $\theta_{tilt}$ is the tilt angle with the horizontal plane.

Supporting Video S6:

Comparative effect of surface tension on the sliding dynamics of droplets

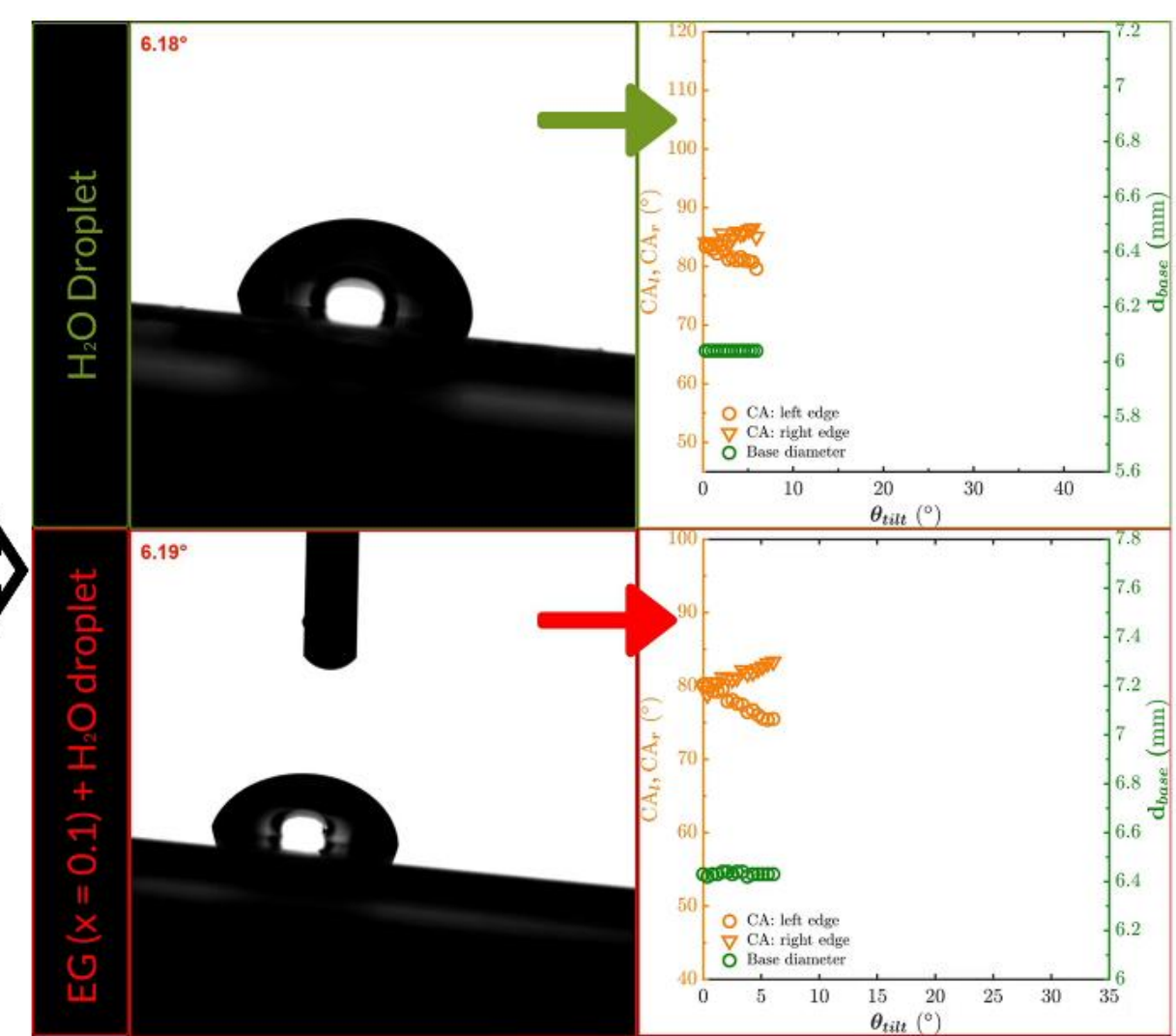

**Supporting Video S7:** Comparative effect of surface tension on the sliding dynamics in the presence of microparticles within the droplets

$CA_l$, $CA_r$ represent the contact angles at the trailing and leading edges, while $d_{base}$ denotes the projected contact diameter of the droplet. $\theta_{tilt}$ is the tilt angle with the horizontal plane.

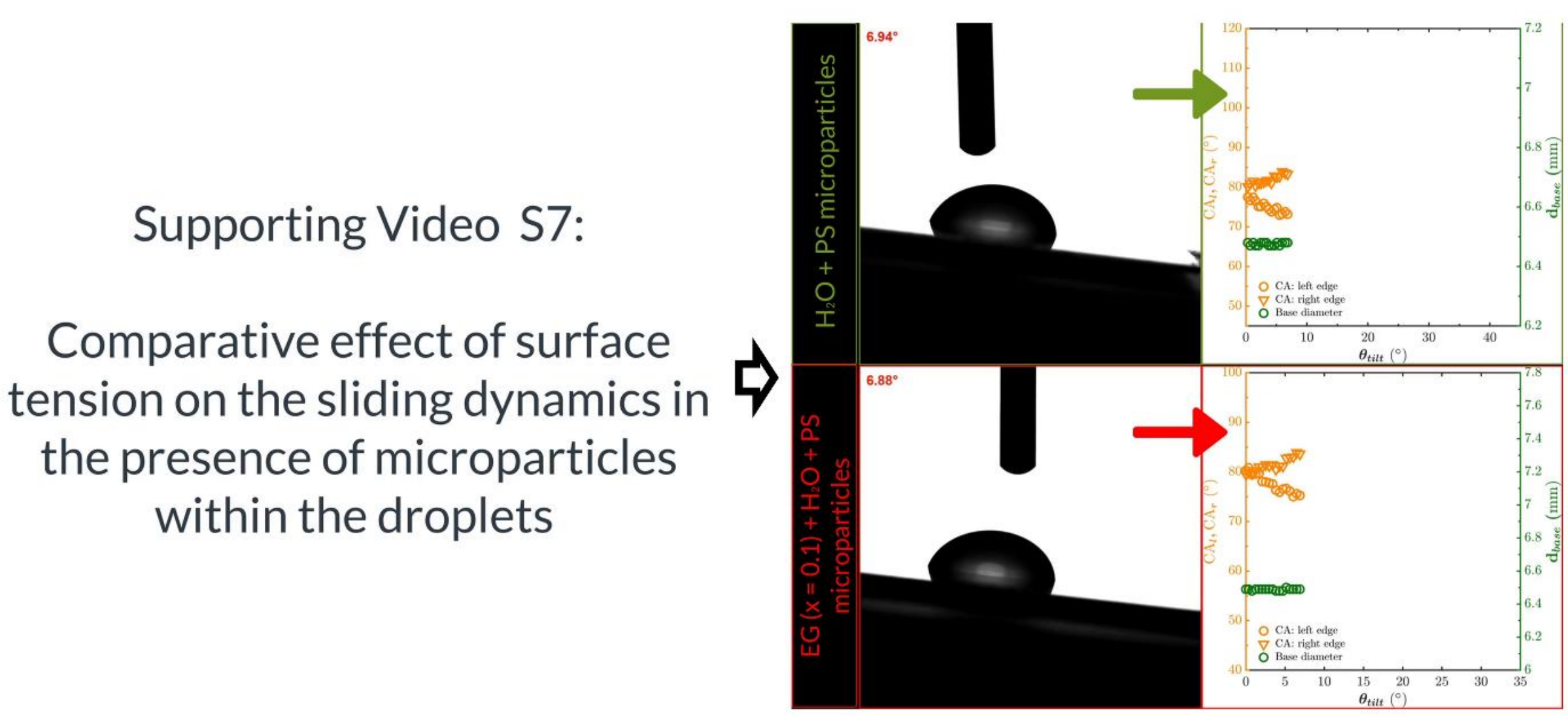

## 2. Supporting Figures

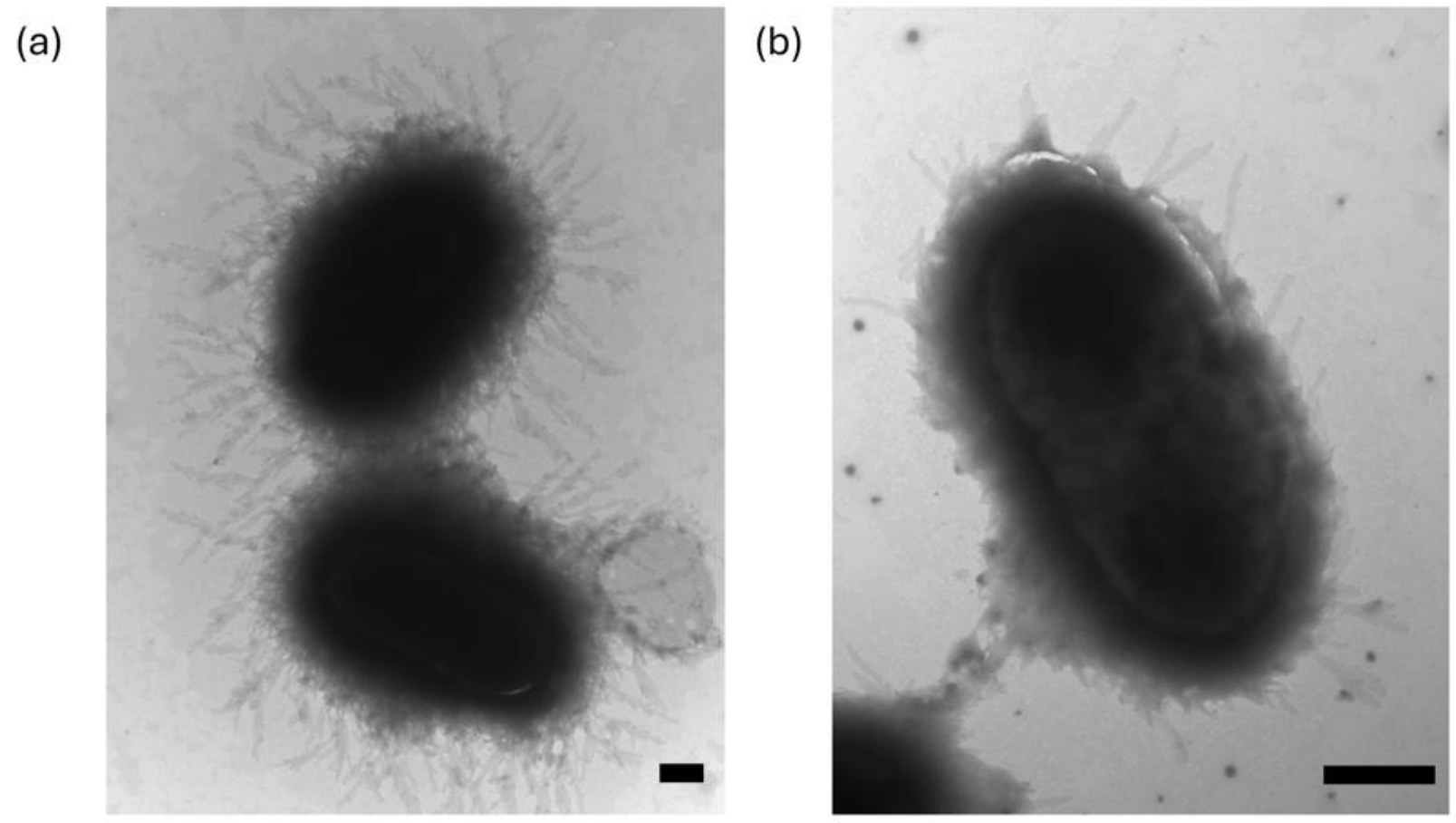

**Figure S1: Transmission electron microscopy (TEM) images of the stained bacteria samples at two different magnifications, which clearly show the presence of flagella responsible for the motility.** 1% (wt/v) ammonium molybdate is used for staining. Scale bar 500 nm.

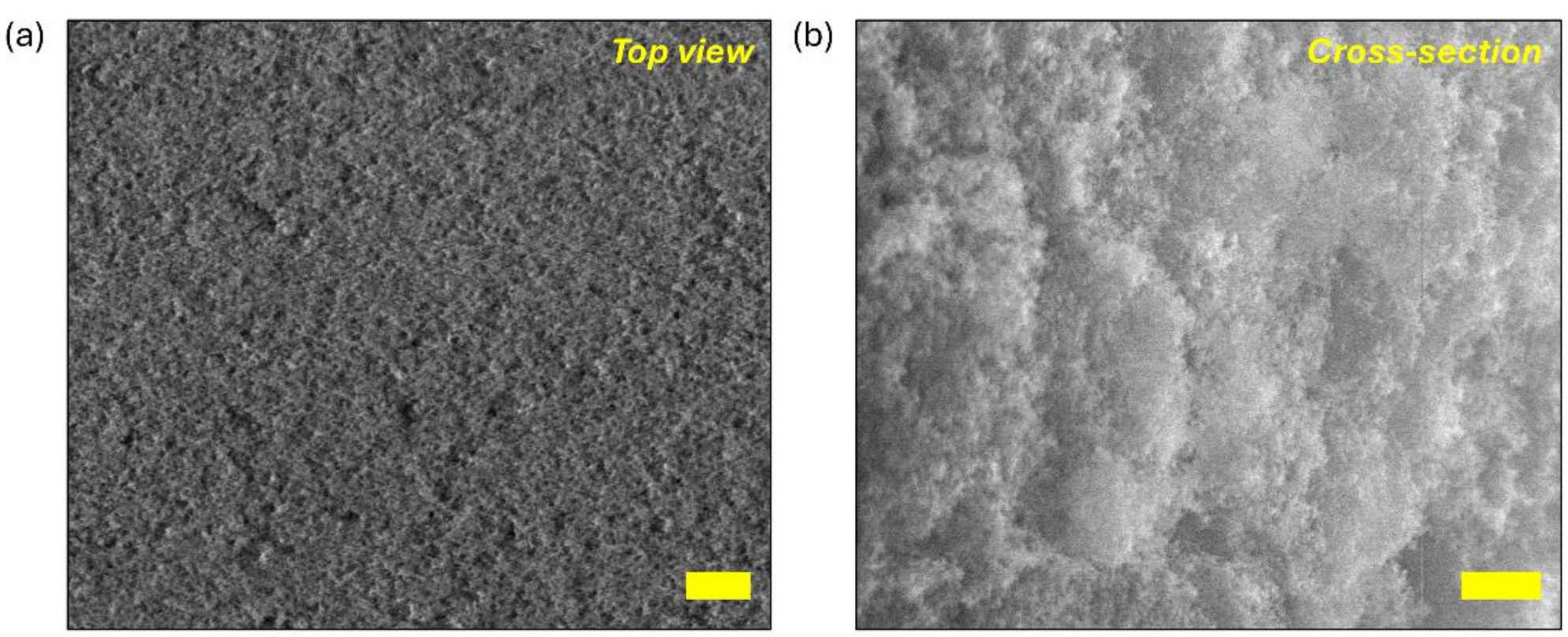

**Figure S2: Scanning electron microscopy (SEM) of the superhydrophobic NeverWet surface.** (a) shows Top view, while (b) presents the cross-sectional view. Coating thickness ~ 45 µm. Scale bar 1 µm

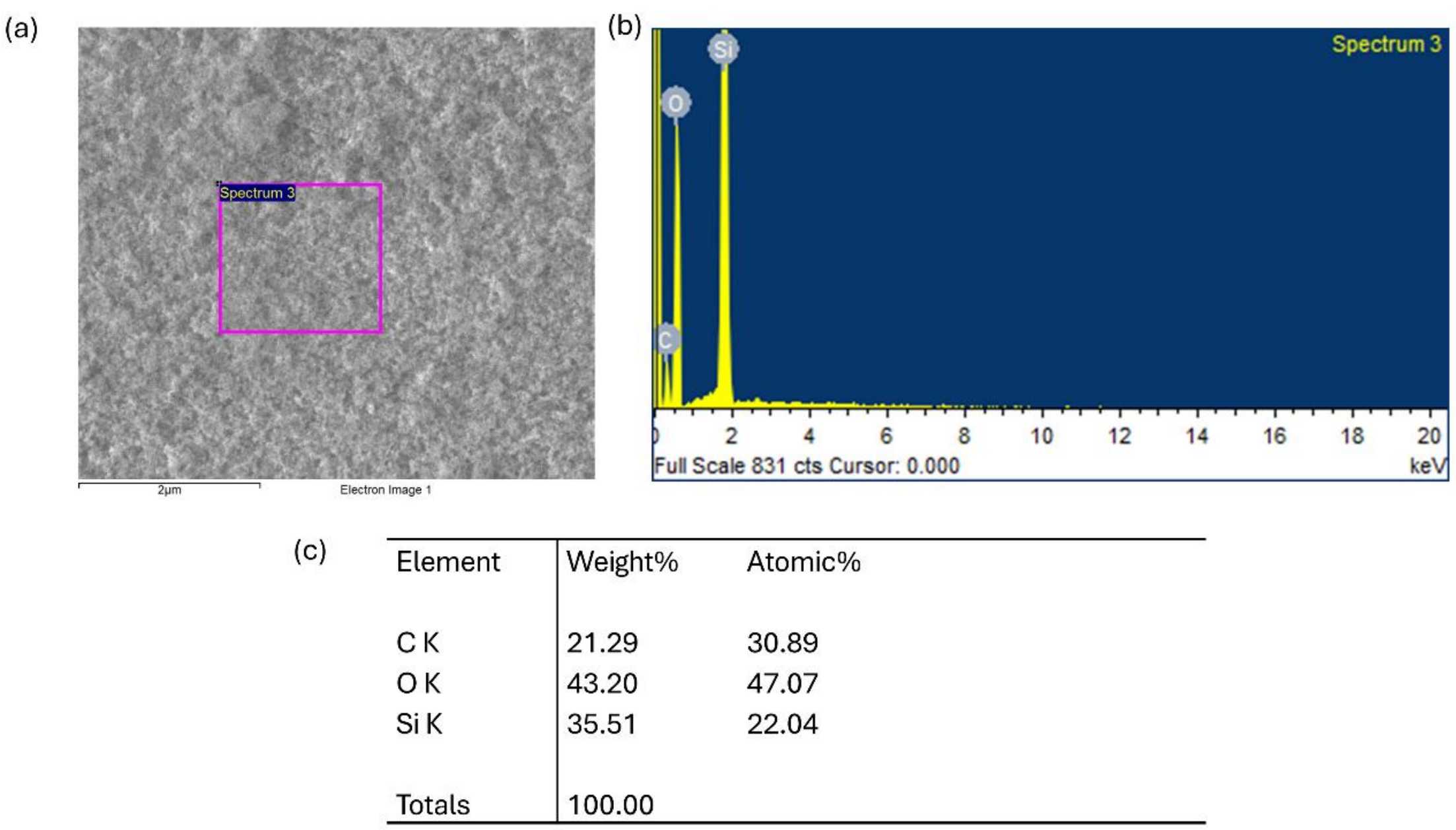

| Element | Weight% | Atomic% |
| --- | --- | --- |
| C K | 21.29 | 30.89 |
| O K | 43.20 | 47.07 |
| Si K | 35.51 | 22.04 |
| Totals | 100.00 | |

**Figure S3: Energy Dispersive X-ray (EDX) spectroscopy of the superhydrophobic NeverWet sample.** (a) shows the SEM image with the maroon square indicating the region of interest for EDX analysis. (b) shows the EDX spectra, while (c) reports the results of the elemental analysis from the EDX spectra.

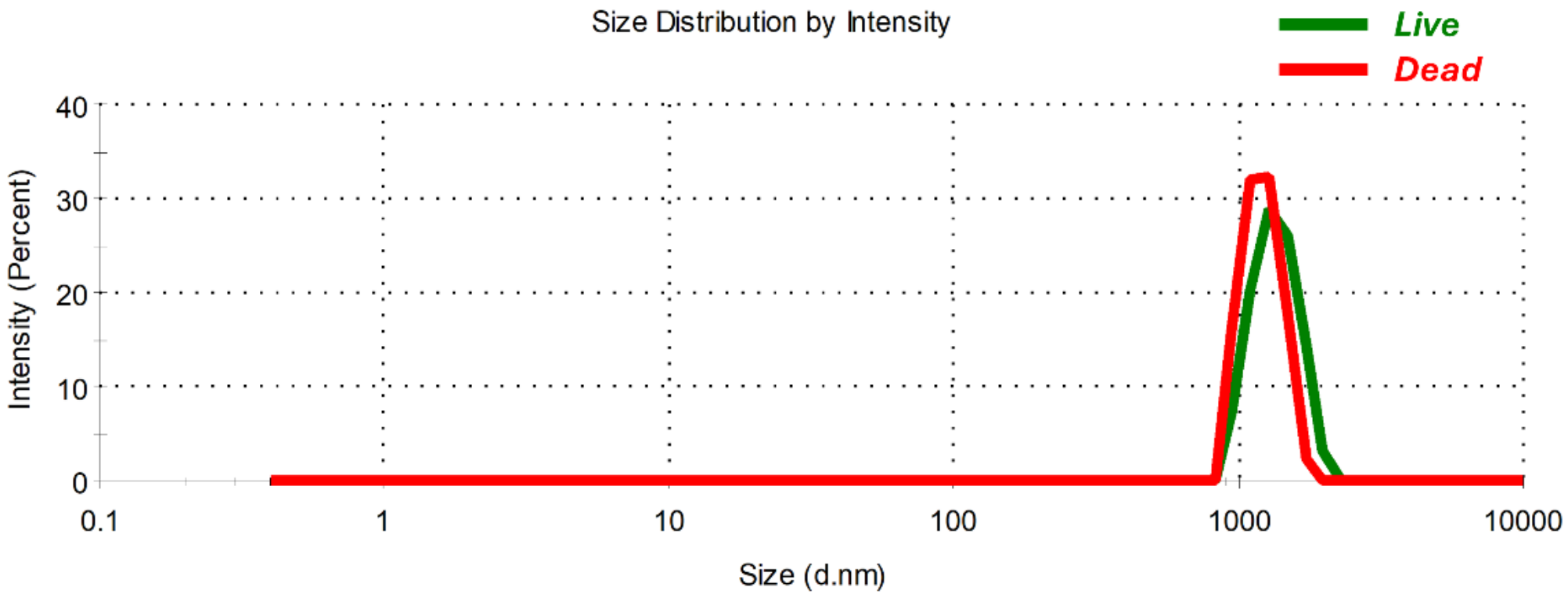

**Figure S4: Comparative size-distribution analysis between live and dead bacteria using Dynamic Light Scattering (DLS).** The sizes corresponding to the intensity peaks are similar (1360±247.4 nm for live vs. 1219±184.9 nm for dead bacteria). The dead sample has a high polydispersity index, indicating some possibility of aggregate formation from the lysis products.

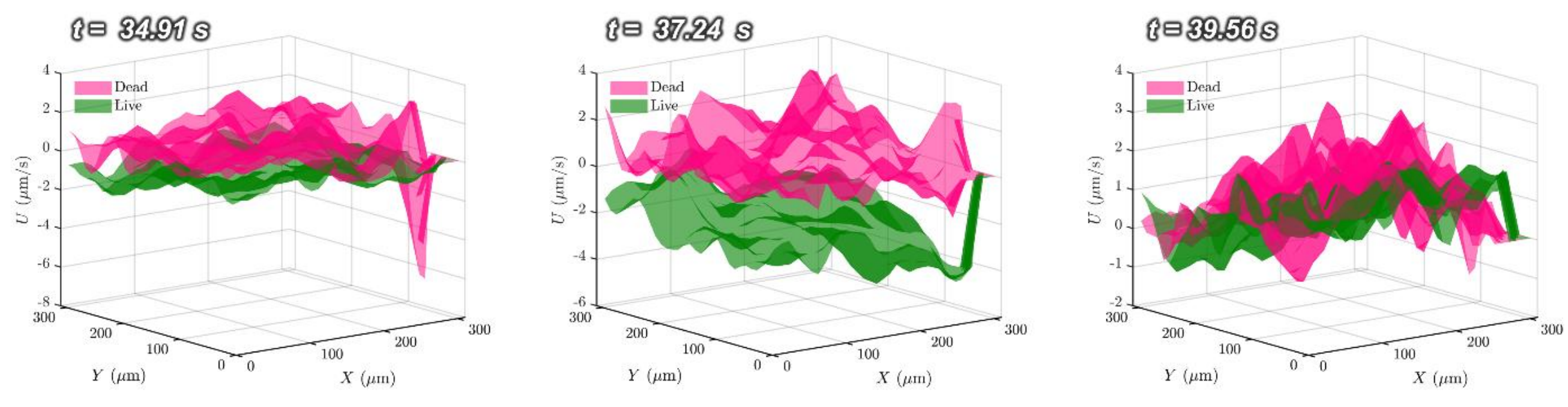

**Figure S5. Direct visual comparison of time-resolved $U$-velocity surface plots for live (green) and dead (magenta) bacterial droplets, plotted on shared axes for each time point**. The data correspond to the same timestamps as shown in Figure 4(a), enabling direct magnitude comparison across live and dead conditions. Each subplot pair highlights the systematically higher positive $U$-velocity observed in dead samples, indicative of consistent motion toward the contact line.

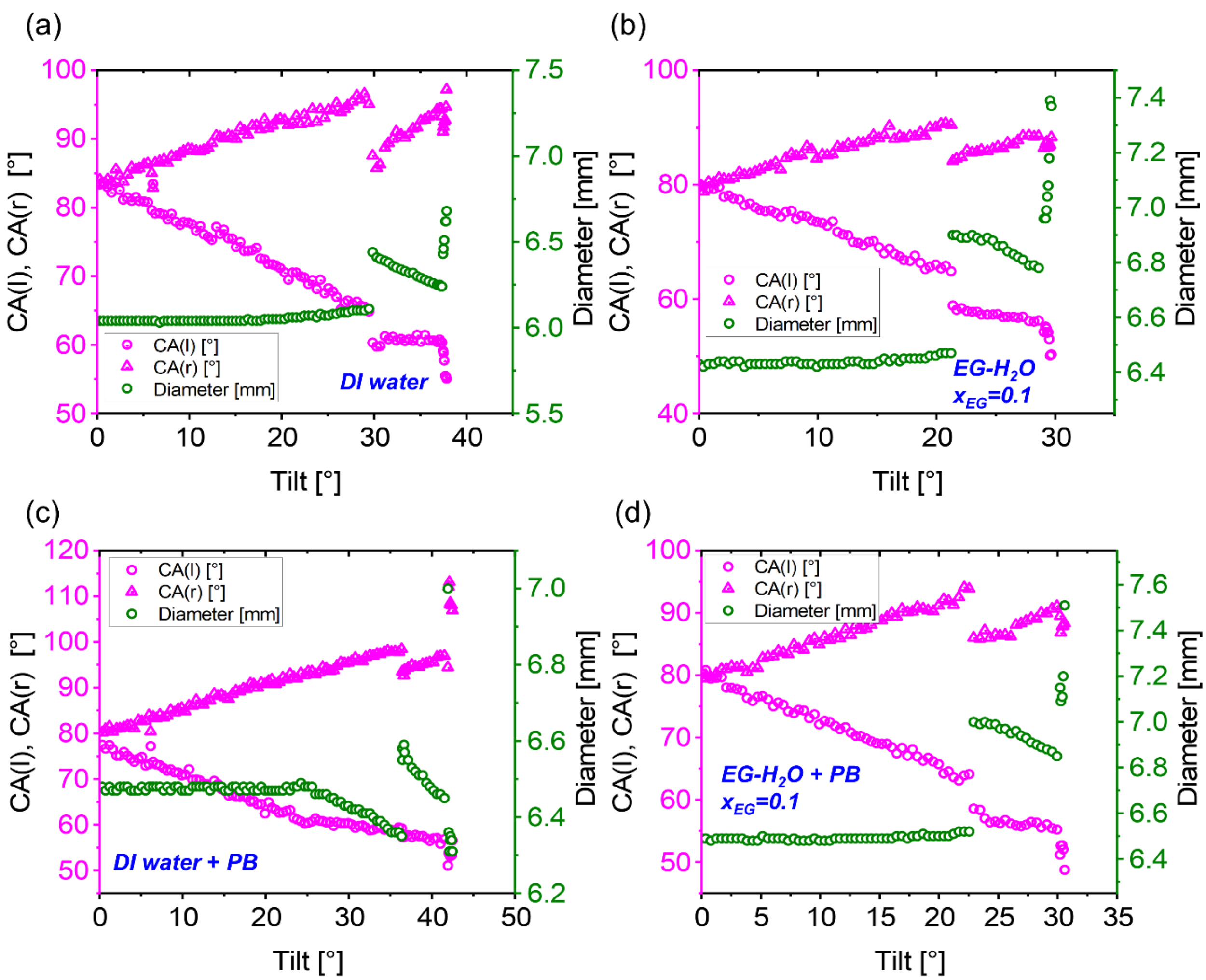

**Figure S6: Tilt-induced sliding behavior of proxy suspensions on a moderate wettability PMMA surface.** (a) Pure DI water, (b) binary mixture of DI water and ethylene glycol (EG) (mole fraction, $x_{EG} = 0.1$), (c) DI water doped with 1 μm-sized polystyrene beads (concentration ≈ $10^7$ particles/mL), and (d) DI water–ethylene glycol mixture ($x_{EG} = 0.1$) doped with polystyrene beads. The evolution of the contact angles of the leading-edge (CA(r)) and the trailing edge (CA(l)), and projected (contact) diameter as a function of tilt angle is shown for each case.